%% file: LaR15-StatComp.tex
\RequirePackage{fix-cm}
\documentclass[smallextended]{svjour3}       
\smartqed  
\usepackage{graphicx}
\usepackage{amsfonts,amsmath,amssymb,epsfig,epsf,psfrag}
\usepackage{color}
\usepackage{xcolor,xspace}
\RequirePackage{natbib}

\newcommand{\abf}{{\bf a}}
\newcommand{\alphabf}{\mbox{\mathversion{bold}{$\alpha$}}}
\newcommand{\Bcal}{\mathcal{B}}
\newcommand{\Beta}{\mbox{Beta}}
\newcommand{\cbf}{{\bf c}}
\newcommand{\dd}{\mbox{d}}
\newcommand{\Esp}{\mathbb{E}}
\newcommand{\Dir}{\mbox{Dir}}
\newcommand{\Ibb}{\mathbb{I}}
\newcommand{\Jcal}{\mathcal{J}}
\newcommand{\mbf}{{\bf m}}
\newcommand{\Mcal}{\mathcal{M}}
\newcommand{\pibf}{\mbox{\mathversion{bold}{$\pi$}}}
\newcommand{\thetabf}{\mbox{\mathversion{bold}{$\theta$}}}
\newcommand{\btheta}{\boldsymbol{\theta}}
\newcommand{\Xbf}{{\bf X}}

\newcommand{\Zbf}{{\bf Z}}

\newcommand{\proofend}{$\blacksquare$\bigskip}

\begin{document}

\title{Variational Bayes Model Averaging for  Graphon Functions and Motif Frequencies Inference in $W$-graph Models\footnote{The final publication is available at Springer via http://dx.doi.org/[10.1007/s11222-015-9607-0]}
}
\subtitle{}

\titlerunning{Variational Bayes Model Averaging for $W$-graph Inference}        

\author{Pierre Latouche         \and
        St\'ephane Robin 
}


\institute{P. Latouche \at
              Laboratoire SAMM, EA 4543,    Universit\'{e}   Paris   1 Panth\'{e}on-Sorbonne, {\sc France} \\
              \email{pierre.latouche@univ-paris1.fr}           
           \and
           S. Robin \at
              AgroParisTech, UMR 518 MIA, Paris, {\sc France} \\
		    INRA, UMR 518 MIA, Paris, {\sc France}
}

\date{Received: date / Accepted: date}

\maketitle

\begin{abstract}
$W$-graph refers to a general class of random graph models that can be
seen as a random graph limit.  It is characterized by both
its graphon function and its motif frequencies.  In this paper, 
{ relying on an existing variational Bayes algorithm for the stochastic block models along with the corresponding weights for model averaging, we derive an estimate of the graphon function as an average of stochastic block models with increasing number of blocks}. In  the same
framework,  we  derive  the  variational posterior  frequency  of  any
motif.  A simulation  study and  an illustration  on a  social network
complete our work.
\end{abstract}

\input{Intro.tex}


\input{InferenceW.tex}

\input{Motifs.tex}

\input{Simuls.tex}


\input{Examples.tex}


\input{Conclusion.tex}

\newpage
\appendix
\section{Appendix}


\input{App-InfW.tex}
\input{App-Motifs.tex}


\begin{acknowledgements}
The   authors  thanks   Stevenn  Volant   for  helpful   comments  and
discussions. The authors also thank the anonymous reviewer for his helpful remarks on our work.
\end{acknowledgements}

\bibliography{biblio}
\bibliographystyle{astats}


\end{document}

%% file: Intro.tex
\section{Introduction}

\paragraph{$W$-graph.}
The $W$-graph model has been intensively studied in the probability literature. From a theoretical point of view, it defines a limit for dense graphs \citep{LoS06}, but it can also be casted into a general class of inhomogeneous random graph models \citep{BJR07} involving some hidden latent space. A $W$-graph is characterized by the so-called 'graphon' function $W$, where $W(u, v)$ is the probability for two nodes with respective latent coordinates $u$ and $v$ (both taken in $[0, 1]$) to be connected. The precise definition of a {$W$-graph model} is given at the end of this section. Because of very weak assumptions about the graphon function, the $W$-graph model is very flexible and can result in a large variety of network topologies. 

\begin{figure}[ht]
  \begin{tabular}{ccc}
    \includegraphics[width=.3\textwidth]{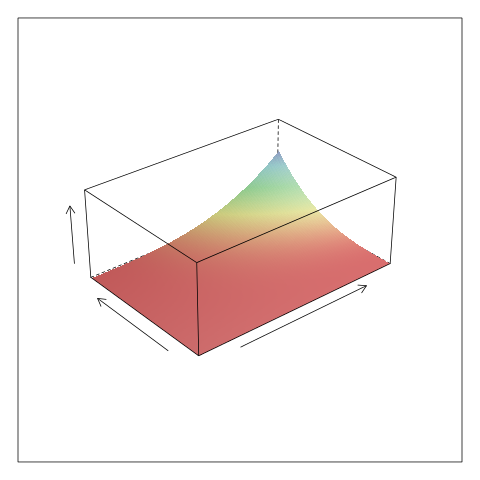} &
    \includegraphics[width=.3\textwidth]{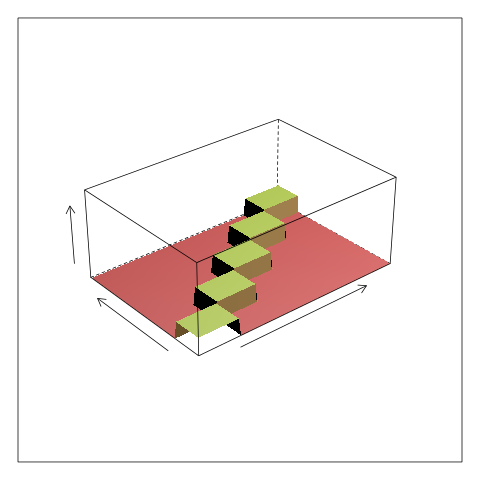} &
    \includegraphics[width=.3\textwidth]{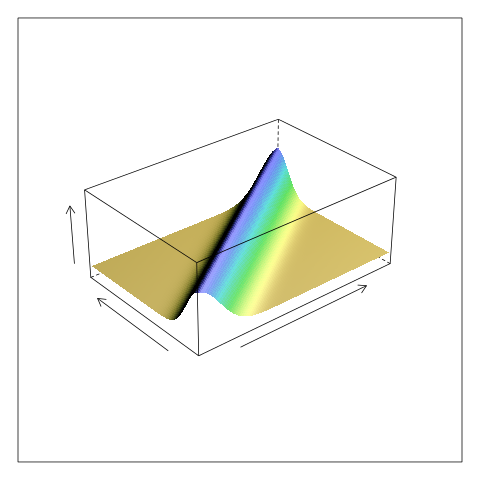} 
  \end{tabular}
  \caption{Graphon function for some typical random graph models. Left: scale-free network. Middle: community network. Right: small world network. \label{fig:ex.graphon}}
\end{figure}

The $W$-graph model suffers an identifiability issue as, for any
measure-preserving transformation $\sigma$ of  $[0, 1]$ into $[0, 1]$,
the  graphon  function  $W_\sigma(u,  v)  =  W(\sigma(u),  \sigma(v))$
results in the same $W$-graph model as {with} the function $W$. This issue  is often  circumvented by further  assuming that  the mean density  $\int  W(u,  v) \dd  v$  is  an  increasing function  of  $u$
\citep{proceedingsbickel2009}. 
However, \cite{DiJ08} showed that subgraphs (called hereafter motifs) frequencies are invariant and constitute intrinsic characteristics of a $W$-graph. 
\paragraph{Interpreting of the graphon.} 
The graphon function provides a two-dimensional representation of the global topology of the network, without any prior assumption as for the form of the degree distribution, or the existence of clusters in the graph. It is the limiting adjacency matrix of the network. For more details, we refer to the work of Lov\'{a}sz and his coauthors (see for instance \cite{LoS06}). We emphasize that the connection between the Aldous-Hoover theorem, which is an extension of deFinetti's theorem to exchangeable arrays, and the notion of graph limits, was made by \cite{DiJ08}. As shown in Section \ref{Sec:Experiments}, the graphon function can help in understanding the organization of the network and offers an alternative visualization that is especially useful for large graphs. \\
Figure \ref{fig:ex.graphon} provides some examples of graphon functions. A scale free network \citep{BaA99} is highly concentrated around a small fraction of nodes with high degree. Such a degree distribution can be retrieved using a graphon similar to this used in the simulation study (see Section \ref{Sec:Simuls}). The concentration around the central nodes is revealed by the peak in the upper right corner of graphon surface. A community network {\citep{girvan2002community}, where nodes tend to connect to nodes of the same community,} is characterized by a block-diagonal structure. A small-world network \citep{Was98,BaR06} is defined by a majority of connexions between neighboring nodes revealed by high values of the graphon function along the diagonal. Edges between non-neighbor  nodes, which provide the 'small world' property, are made possible by the non-zero value of the graphon function apart from the diagonal. {Thus the} graphon function summarizes the global topology of the network.  \\ 
On the other hand, the characterization in terms of motifs provides an information about the local organization of the network. Such a characterization has been used to depict the organization and the functioning of biological networks \citep{MSI02}. Because the topology of a $W$-graph only depends on the respective latent location of pairs of nodes, the empirical frequency of motifs of size larger that three can be used to assess the goodness-of-fit of the model.

\paragraph{Statistical inference.}
Until recently, little  attention has been paid to  the statistical
inference of a  $W$-graph model, based on an  observed network. The earliest reference on graphon estimation is \cite{kallenberg99} who showed the weak convergence of a function, which can be seen as an empirical graphon, to the graphon function. A general framework was considered without further modeling assumptions. Contrary to this work, \cite{PLV10} derived a parametric exchangeable model along with an MCMC-like algorithm to perform inference on real data, resulting in a heavy computational burden. An alternative parametric approach was proposed earlier by \cite{hoff2008}. The graphon estimation as a nonparametric problem was first formulated in \cite{lloyd2012}. The authors relied on Gaussian process priors to build the nonparametric scheme and also on an MCMC-like algorithm for the inference. Since then, many approaches have been proposed, in parallel to the  present work \citep{ACC13,WoO13,asta14,Cha12,Borgs2015}. We emphasize that a series of methods has considered total variation estimation \citep{chan2014} and the corresponding two or three step inference procedures \citep{Yang2014}. \\
\cite{WoO13} developed a theoretical framework   for   the  non-parametric   estimation   of  the   graphon function. This latter approach also relies on the connection between the SBM  and $W$-graph models.  However, while they focus on the blockwise constant approximation of  the graphon  function, we rely  in this paper  on  approximate  posterior distributions of  the SBM  model, which results in a smoother estimate of the graphon function.

\paragraph{Link with SBM.}
In parallel to $W$-graph, the stochastic  block model (SBM: \cite{NoS01}) has  been used in a large  variety of domains, from
sociology to biology,  and many efforts have been made  in view of its
inference.   The  proposed  inference   techniques  range   from  MCMC
\citep{NoS01} to a degree based algorithm \citep{CDR12},
including variational expectation maximization (VEM)  \citep{DPR08} and variational Bayes EM (VBEM) \citep{articlelatouche2012}. 
SBM states that each node belongs to a certain class (in finite number) and that the probability for two nodes to be connected depends on the class they belong to. As shown in the next section, SBM corresponds to a $W$-graph for which the graphon function is blockwise constant. Interestingly, the frequency of motifs in SBM has been studied by \cite{PDK08}, who provided explicit formulas.

\paragraph{Contribution.}
Our purpose in this paper is to rely on some of the statistical works developed for the SBM model in order to carry out the inference of the $W$-graph model. Considering SBM as a proxy of the $W$-graph model, we propose a complete inference procedure for both the graphon function and the frequency of any motif. The method is based on the variational Bayes approach proposed by \cite{articlelatouche2012}. \\
The SBM postulates that nodes can belong to a certain number $Q$ of classes, which does not make sense for a $W$-graph. We show how the posterior distributions conditional to the number of clusters can be integrated out in order to provide an estimation of the posterior distribution of  the graphon  function. In practice,  this integration leads to a smooth version of the blockwise constant approximation. Indeed, this property is desirable when the monotone version of the graphon is expected to be smooth. \\
In the same spirit, we provide a variational Bayes estimate of the frequency of any network motif (or sub-graph). These frequencies allow us to search for unexpectedly frequent motifs in the network and we suggest to use these results to assess the goodness-of-fit of the model.

\paragraph{Outline.}
The paper is organized as follows. In Section \ref{Sec:InferenceW}, we
present  the  connexion  between  SBM  and $W$-graph,  we  remind  the
principle of  the variational Bayes method and  derive the approximate
posterior  distribution of the  graphon function.  We follow  the same
line in  Section \ref{Sec:Motifs} to derive  the approximate posterior
mean  of motifs  frequencies.  The performances  of  the approach  are
studied via  simulation in  Section \ref{Sec:Simuls} and  the proposed
method is applied to a subset of the French political blogosphere network,
in Section \ref{Sec:Experiments}.

\paragraph{Notations and definition of the $W$ random graph.} 
All along the paper, we will use the following notations for the $W$-graph model. We consider $n$ nodes labeled with index $i = 1, \dots n$. A latent variable $U_i$ drawn uniformly over $[0, 1]$ is associated with each node $i$, the $U_i$'s being mutually independent. The edges $\{X_{ij}\}_{i < j}$ are then drawn independently conditionally on the latent $\{U_i\}$ as $X_{ij} | U_i, U_j \sim \Bcal[W(U_i, U_j)]$, where $W: [0, 1] \times [0, 1] \rightarrow [0, 1]$ denotes the graphon function {and $\Bcal(\cdot)$ is the Bernoulli distribution}.

%% file: InferenceW.tex
\section{Inference of the graphon function} \label{Sec:InferenceW}

We propose to estimate the function $W$ via the inference of a
stochastic block model. The aim of this section is to recall previous results on the
variational Bayes inference of SBM and to show how they can be used to
estimate $W$.

\paragraph{Stochastic block model (SBM).}
We first recall the definition of the SBM model \citep{NoS01}. The $n$ nodes are
supposed to be spread into $Q$ groups with proportions $\alphabf =
(\alpha_1, \dots, \alpha_Q)$. More precisely, the nodes are associated
with independent (unobserved) labels $Z_i$ drawn from a multinomial distribution
$\Mcal(1; \alphabf)$. Connections are ruled by a $Q \times Q$
connectivity matrix $\pibf = [\pi_ {q\ell}]$, where $\pi_ {q\ell}$ is
the connection probability between a node from group $q$ and a node
from group $\ell$ ($\pibf$ has to be symmetric for undirected
graphs). The  edges of the graph  are then drawn  independently from a
Bernoulli distribution,
conditionally on the labels $Z_i$, as $X_{ij} | Z_i, Z_j \sim
\Bcal(\pi_{Z_i, Z_j})$. \\ 
In the sequel, we shall denote $\Zbf = \{Z_i\}$ the set of 
unobserved labels, $\Xbf = \{X_{ij}\}$ the set of observed edges
and $\thetabf = (\alphabf, \pibf)$ the set of model parameters.

\paragraph{Connection between SBM and $W$-graph.}
SBM corresponds to the case where $W$ is blockwise constant, with
rectangular blocks of size $\alpha_k \times \alpha_\ell$ and height
$\pi_{q\ell}$. More precisely, denoting the cumulative proportion 
\begin{equation} \label{Eq:CumProp}
  \sigma_q = \sum_{j=1}^q \alpha_j,  
\end{equation}
if we define the binning function
\begin{equation*} 
C_{\alphabf}(u) = 1 + \sum_{q=1}^Q \Ibb\{\sigma_q \leq u\},
\end{equation*}
and if we take
\begin{equation} \label{Eq:PhiHat}
W(u, v) = \pi_{C(u), C(v)}, 
\end{equation}
the resulting $W$-graph model corresponds to the SBM model with parameters $(\alphabf, \pibf)$.

\cite{PLV10} also considered a blockwise constant model; these authors {used} a self similarity transformation to increase the number of {blocks} to gain flexibility keeping the number of parameters small. However, the connection with $W$-graphs is not made explicitly in this article.

\paragraph{Identifiability.}
As pointed out by \cite{proceedingsbickel2009}, the $W$-graph is not
identifiable, as any measure-preserving transformation of the interval
$[0, 1]$ would provide the same random graph. Following these authors,
we fix the version of $W$ to be estimated using the constraint that
the function $D(u) = \int W(u, v) \dd v$ is monotonic increasing. For
consistency, we require the corresponding condition for the SBM to be
fitted, that is: $d_q = \sum_\ell \alpha_\ell \pi_{q\ell}$ increases
with $q$.

\subsection{Variational Bayes inference of SBM}

The inference of SBM has received many attention in the last
decade. Briefly speaking the main pitfalls lies in the determination
of the conditional distribution of the labels in $\Zbf$, given the
observation $\Xbf$, which displays an intricate dependency
structure. Both Monte-Carlo sampling \citep{NoS01} and variational
approximations \citep{DPR08} have been proposed, but the later scale
better. In this paper, we will use the variational Bayes approximation
proposed in \cite{articlelatouche2012}, which provides a closed-form
approximate posterior distribution of the parameters $\thetabf$ and of
the hidden variables in $\Zbf$ (denoted
$\widetilde{p}_{\btheta}(\thetabf)$ and
$\widetilde{p}_{\Zbf}(\Zbf)$). We recall that this approximation is
obtained through the maximization with respect to
$\widetilde{p}_{\btheta}(\cdot)$ and $\widetilde{p}_{\Zbf}(\cdot)$ of the
functional
\begin{equation} \label{Eq:Jfunction}
\Jcal = \log P(\Xbf) - KL(\widetilde{p}_{\btheta}(\thetabf)
\widetilde{p}_{\Zbf}(\Zbf) || P(\thetabf, \Zbf|\Xbf)),
\end{equation}
where $KL$ stands for the K\"ullback-Leibler divergence.  Our estimate
of the function $W$ strongly relies on this approximate posterior
distribution, which has been shown to be reliable by \cite{GDR11}. In
the sequel, we shall use a tilde to mark approximate posterior
variational distributions and probabilities.

The variational Bayes inference of SBM can be achieved using the VBEM
algorithm described in \cite{BeG03}. As SBM can be casted into the
exponential family framework, using conjugate priors for the
parameters
\begin{eqnarray*} 
  \alphabf & \sim & \Dir(\abf^0) \qquad \text{where } \abf^0 =
  (a^0_1, \dots, a^0_Q), \nonumber \\ 
  \pi_{q, \ell} & \sim & \Beta(\eta^0_{q, \ell}, \zeta^0_{q, \ell}),
\end{eqnarray*}
(where $\Dir$ stands for the Dirichlet distribution), the variational Bayes posterior approximation states
that $\alphabf$ and the $\pi_{q\ell}$ are all conditionally
independent given $\Xbf$ with distributions
\begin{eqnarray} \label{Eq:PosteriorTheta}
  \alphabf | \Xbf & \sim & \Dir(\abf) \qquad \text{where } \abf =
  (a_1, \dots, a_Q), \nonumber \\ 
  \pi_{q, \ell} | \Xbf & \sim & \Beta(\eta_{q, \ell}, \zeta_{q, \ell}).
\end{eqnarray}
The expressions of the $a_q$, $\eta_{q\ell}$ and $\zeta_{q\ell}$ as functions of $a^0_q$,  $\eta^0_{q\ell}$, $\zeta^0_{q\ell}$ and $\Xbf$ can
be found in \cite{articlelatouche2012}.

No general  guaranty exists about the  theoretical properties of
  variational (Bayes)  estimates. Still, SBM  appears to be  a special
  case  were the  consistency of  the variational  estimates has  been
  shown in a frequentist  setting (\cite{CDP12}, \cite{MAM15}). As for
  the variational Bayes estimates, the simulation study carried out by
  \cite{GDR11} shows  that the  approximate posterior  distribution is
  accurate even for networks with only few tens of nodes.

\subsection{Posterior distribution of the function $W$}\label{ssec:postW}

We now derive the approximate posterior distribution of the function
$W$, at given coordinate $(u,  v)$. To this aim, \eqref{Eq:PhiHat} has
to  be   integrated  with  respect  to   the  (approximate)  posterior
distributions of both
$\pibf$ and $\alphabf$. 
\begin{proposition} \label{Prop:PosteriorW}
  For given $(u, v) \in [0, 1]^2$, $u \leq v$, using a SBM with $Q$
  groups, the variational Bayes approximate pdf of $W(u, v)$ is
  $\widetilde{p}(w(u, v)|\Xbf, Q)$ can be computed exactly  with complexity $O(Q^2)$.
\end{proposition}

{The key point is that the cumulative distribution function (cdf) of a Dirichlet distribution can be calculated via simple recursions given in \cite{GoS10}. The rest of the proof relies on standard algebraic manipulations and is postponed to Appendix \ref{App:InferenceW}. }

\bigskip
The approximate posterior mean comes as a direct by-product of
Proposition \ref{Prop:PosteriorW}: $\widetilde{\Esp}[W(u, v) | \Xbf] =
$
$$ 
\sum_{q \leq \ell} \frac{\eta_{q, \ell}}{\eta_{q, \ell} + \zeta_{q,
    \ell}} \left[F_{{q-1}, {\ell-1}}(u, v; \abf) - F_{{q},
    {\ell-1}}(u, v; \abf) - F_{{q-1}, {\ell}}(u, v; \abf) + F_{{q},
    {\ell}}(u, v; \abf)\right]{,}
$$ 
{where $F_{q, \ell}(u, v; \abf)$ denotes the joint cdf of $(\sigma_q,
\sigma_\ell)$, as defined in \eqref{Eq:CumProp}, when $\alphabf$ has a Dirichlet distribution
$\Dir(\abf)$.} The approximate posterior standard deviation can be computed as well.

\paragraph{Variable number of groups.}
Denoting $\widehat{\Jcal}_Q$ the maximum of the function defined in (\ref{Eq:Jfunction}), \cite{articlelatouche2012} derived a close-form expression of $\widehat{\Jcal}_Q$ and showed that it can be used as a model selection criterion, choosing $\widehat{Q} = \arg\max_Q \widehat{\Jcal}_Q$. 
Thus, in straightforward scenarios where the (hidden) $W$-graph model could be casted as a unique SBM, \emph{i.e.} graphon function is exactly blockwise constant, this framework provides a way to estimate the number of blocks as $\widehat{Q}^2$. \\
Still, because SBM is mostly used as a proxy for $W$-graph, it may seem more realistic to assume that no true number of groups $Q$ does actually exist. Therefore we rather consider here a model averaging approach in which the inferred $W$-graph is an average of a series of SBM with increasing number of groups $Q$.
\cite{VMR12} derived the variational Bayes approximation of $p(Q|\Xbf)$ and prove that, if a uniform prior over $Q$ is used, the variational approximation satisfies
$$
\widetilde{p}(Q | \Xbf) \propto \exp \widehat{\Jcal}_Q,
\qquad \sum_Q \widetilde{p}(Q | \Xbf) = 1.
$$
In this case, the variational Bayes approximate posterior distribution
of $W(u ,v)$, integrated over the number of groups, is simply
$$ 
\widetilde{p}(w|\Xbf) = \sum_Q \widetilde{p}(Q | \Xbf) \;
\widetilde{p}(w|\Xbf, Q).
$$ 
We remind that $\widehat{\Jcal}_Q$ is the difference between the
  true marginal likelihood of the data with $Q$ groups, $\log P(X|Q)$ and the $KL$ divergence between the variational approximation of the condition distribution $P(\thetabf, \Zbf|\Xbf, Q)$ and this distribution itself. The regular Bayesian model averaging would directly rely on $P(X|Q)$ to weight each considered model (\cite{HMR99}). Because of the accuracy of the variational Bayes approach for SBM (\cite{GDR11}, the $KL$ divergence is expected to be small, so the variational weights are close to the theoretical ones.

%% file: Motifs.tex
\section{Motif probability} \label{Sec:Motifs}

As recalled above, the $W$-graph model suffers a deep identifiability
problem. However, as shown in \cite{DiJ08}, the distribution of the
number of occurrences of patterns or motifs turns out to be invariant,
and therefore characteristic of a given $W$-graph model. In this section,
we show how variational Bayes inference of SBM can be used to estimate
a key quantity of such a distribution, namely the occurrence
probability of the motif.

The number of occurrences of a given motif in random graphs has been
intensively studied in Erd\"os-R\'enyi graphs 
\citep[see for instance][]{Sta01} and some results about $W$-graphs can be found
in \cite{DiJ08} and \cite{BJR07}.  However, the exact distribution
for an arbitrary motif can not be determined in general. On the other
hand, \cite{PDK08} derived a general approach to derive the moments of
the number of occurrences in stationary graphs. These moments only
depend on the size of the graph, on the number of automorphisms of the
motif and on the occurrence probability of the motif (and of its
super-motifs).

The occurrence probability of a motif is therefore a key quantity to
characterize the number of occurrences of a motif in a random
graph. In this section, we recall the definition of a motif occurrence
and of the occurrence probability. Then, we show how variational Bayes
inference of SBM can be used to estimate this occurrence probability in
a $W$-graph.

\subsection{Motif probability}

\paragraph{Definition of a motif.}
A motif can be defined as a sub-graph with prescribed edges. More
precisely, a motif with size $k$ is completely defined by the $k
\times k$ 0-1 adjacency matrix $\mbf$, where $m_{a, b} = 1$ if there
is an edge between node $a$ and $b$, 0 otherwise. Figure
\ref{Fig:Motifs}, given in \ref{App:Motifs}, displays some typical motifs
and their corresponding adjacency matrices $\mbf$.

As for the occurrence of a motif, we use here the definition used in
\cite{PDK08}, which defines an occurrence of $\mbf$ as a set of $k$ nodes
in the graph, such that all edges prescribed in $\mbf$ actually
occur. Formally, we  consider the occurrence indicator of  the motif $\mbf$
at position $\beta=(i_1, \dots, i_k)$, with $i_{1}<\dots<i_{k}$, as
\[
Y_{\beta}(\mbf) = \prod_{1 \leq a < b \leq k} (X_{i_a, i_b})^{m_{ab}}.
\]
In a $W$-graph model, as in all stationary graphs, a given motif $\mbf$ has the
same probability to occur at any position. This probability is called
the occurrence probability of the motif $\mbf$ and we denote it by
$\mu(\mbf)$
\[
\mu(\mbf) = \Pr\{Y_{\beta}(\mbf) = 1\}.
\]

\paragraph{Motif probability in $W$-graph.}
Because the edges are independent conditionally to the latent labels, the probability of a motif $\mbf$ in a $W$-graph has the following general form
$$
\mu(\mbf) = \int \dots \int \prod_{1 \leq a < b \leq k} \left[W(u_a, u_b)\right]^{m_{ab}} \dd u_1 \dots \dd u_k.
$$
We provide a close form version of this result in a special case that will be used in the simulation study.

\begin{proposition} \label{Prop:ProbMotifWProd}
 If the $W$ function has a symmetric product form $W(u, v) = g(u) g(v)$, then
 $$
 \mu(\mbf) = \prod_{1 \leq a \leq k} \xi_{m_{a+}}, 
 \qquad \text{where} \quad
 \xi_{h} = \int g(z)^h \dd z,
 $$
 and $m_{a+} = \sum_{1 \leq b \leq k} m_{ab}$ denotes the degree of vertex $a$ in the motif $\mbf$.
\end{proposition}
The proof is given in Appendix \ref{App:Motifs}.

\subsection{Occurrence probability estimate}\label{ssection:motifprob}

As shown in \cite{PDK08}, for a SBM model, with fixed number $Q$ of groups and with parameters
$(\alphabf,\pibf)$, the form of $\mu(\mbf)$ is given by
\begin{equation} \label{Eq:MumSBM}
\mu(m | \alphabf, \pibf) = \sum_\cbf  \prod_{1 \leq a \leq k} \alpha_{c_a}
 \prod_{1 \leq a < b \leq k} \pi_{c_a, c_b}^{m_{ab}}, 
\end{equation}
where $\cbf$ stands for the labeling of the $k$ nodes: $\cbf = (c_1,
\dots c_k)$, each label $c_a$ being taken in $\{1, \dots Q\}$. 
Keeping $Q$ fixed, but integrating the uncertainty over $\alphabf$ and
$\pibf$, we derive the approximate posterior mean
\begin{equation*}
  \widetilde{\Esp}[\mu(\mbf) |  \Xbf, Q] =\int \int  \mu(m | \alphabf,
  \pibf) \widetilde{p}(\alphabf,\pibf|\Xbf,Q)d\alphabf d\pibf.
\end{equation*}
\begin{proposition} \label{Prop:PosterioMumSBM}
  Using the same notation as in Proposition \ref{Prop:PosteriorW}, the
  approximate variational Bayes posterior mean of the occurrence
  probability under SBM with $Q$ groups is
  \begin{eqnarray*}
    \widetilde{\Esp}[\mu(\mbf) | \Xbf, Q] 
    & = & \left\{
    \left[ 
      \prod_{q \leq \ell}^{Q} \frac{\Gamma(\eta_{q\ell}+\zeta_{q\ell})}{\Gamma(\eta_{q\ell})} 
      \right]
    \frac{\Gamma(\sum_{q=1}^{Q}n_{q})}{\prod_{q=1}^{Q}\Gamma(n_{q})}
    \right\} \\ 
    & & \times \left\{ 
    \sum_{\cbf} \left[ 
      \prod_{q \leq \ell}^{Q}\frac{\Gamma(\eta_{q\ell} +
        \eta_{q\ell}^{\cbf})}{\Gamma(\eta_{q\ell} + \eta_{q\ell}^{\cbf} +
        \zeta_{q\ell})}
      \right] 
    \frac{\prod_{q=1}^{Q}\Gamma(n_{q}+n_{q}^{\cbf})}{\Gamma \left[
        \sum_{q=1}^{Q} (n_{q}+n_{q}^{\cbf}) \right] }\right\},
  \end{eqnarray*}
  where $\cbf = (c_1, \dots, c_k)$, $n_{q}^{\cbf} = \sum_a \Ibb\{c_a =
  q\}$, $\eta_{q\ell}^{\cbf} = \sum_{1 \leq a \neq b \leq k}
  \Ibb\{c_a = q\}\Ibb\{c_b = \ell\} m_{ab}$ for $q \neq \ell$,
  $\eta_{qq}^{\cbf} = \sum_{1 \leq a < b \leq k} \Ibb\{c_a =
  q\}\Ibb\{c_b = q\} m_{ab}$ and $\Gamma(\cdot)$ is the gamma function.
\end{proposition}
The proof is based on the exact calculation of the mean of the
occurrence probability \eqref{Eq:MumSBM} using the variational Bayes
posterior \eqref{Eq:PosteriorTheta} and is postponed to Appendix
\ref{App:Motifs}.

Therefore, integrating $ \widetilde{\Esp}[\mu(\mbf)  | \Xbf, Q] $ over
the number $Q$ of groups, as in Section \ref{ssec:postW}, leads to the
following approximate variational Bayes posterior mean of the
occurrence probability of any motif $\mbf$
\begin{equation}\label{eq:motifBMA}
\widetilde{\Esp}[\mu(\mbf)|\Xbf]=\sum_{Q}\widetilde{p}(Q|\Xbf)
\widetilde{\Esp}[\mu(\mbf) | \Xbf, Q]. 
\end{equation}

\subsection{Testing unexpectedly frequent motifs}\label{ssection:testing}

In  the  following, we  emphasize  that  (\ref{eq:motifBMA}) can  help
in characterizing the count of a motif in a network.  Let us consider
\begin{equation*}
 I_{k}=\left\{\{i_1,\dots,i_{k}\}\subset     \{1,\dots,n\}|i_j    \neq
   i_l,\forall j\neq l\right\},
\end{equation*}
 the  set   of  all   potential   positions  of   $\mbf$  in   the
graph. Permuting the rows as well as the columns of the adjacency matrix $\mbf$
can   lead  to   the  same   motif,  at   each  position   $\beta  \in
I_{k}$. Therefore, denoting $\mathcal{R}(\mbf)$ the set of non
redundant permutations, the count of a motif is defined as 
\begin{equation*}
N(\mbf)=\sum_{\beta \in I_{k}}\sum_{\mbf'\in \mathcal{R}(\mbf)}Y_{\beta}(\mbf').
\end{equation*}
Since the $W$-graph  model is a stationary model,  the expectation and
variance of $N(\mbf)$ have  analytical forms, as shown in \cite{PDK08}
for   general  class   of  stationary   random  graph   models.  While
the calculation of  $\Esp[N(\mbf)]$ is straightforward, the derivation
of $\mathbb{V}(N(\mbf))$  is more technical  and involves super-motifs
which are  made of overlaps between occurrences  of $\mbf$. Therefore,
for the  sake of  the discussion, these  two quantities are  not given
here and we refer to \cite{PDK08}. 
  A key point is that both $\Esp[N(\mbf)]$ and
$\mathbb{V}(N(\mbf))$  involve  occurrence  probabilities  for  which  we
provide estimators (\ref{eq:motifBMA}).  Therefore we
propose  to  replace the  $\mu(\cdot)$  terms  in $\Esp[N(\mbf)]$  and
$\mathbb{V}(N(\mbf))$     with     their     corresponding     estimators
$\widetilde{\Esp}[\mu(\cdot)|\Xbf]$.    Note  that   an
  alternative  approach   consists  in  approximating   the  occurrence
  probabilities themselves using plug-in estimators. We refer to \cite{bickel11} who studied the
asymptotic normality of such plug-in estimates and to \cite{BhB15} who considered a resampling-based approach to estimate the variance of the count.

%% file: Simuls.tex
\section{Simulation study} \label{Sec:Simuls}

We  designed  a  simulation  study   to  assess  the  quality  of  the
variational  Bayes inference  we  propose. Our  study  focuses on  the
estimation  of  both  the  graphon and  the  motifs  frequencies.  The
methodology obviously  depends on the  choice of prior  parameters for
the prior  distributions. In  practice, we set  $a_{q}^{0}=1,\forall q$
and  $\eta_{q,\ell}^{0}=\zeta_{q,\ell}^{0}=1,\forall  (q,\ell)$.  Such
choices induce uniform prior distributions over all model parameters.

\subsection{Simulation design}

\paragraph{Simulation model.}
We considered $W$-graph models with graphon function  $W(u, v) = g(u) g(v)$ where 
\begin{equation} \label{Eq:ModSim}
g(u) = \sqrt{\rho} \lambda u^{\lambda-1}.
\end{equation}
The parameter $\rho$  controls the density of the  graph, meaning that
$\rho$  is the mean  probability for  any two  nodes to  be connected,
while $\lambda$ controls the concentration of the degrees: the higher $\lambda$, the more the edges are concentrated around few nodes. Note that $\lambda=1$ corresponds to the Erd\"os-Rényi model with connection probability $\rho$. Also note that the maximum of $W$ is $\rho \lambda^2$, which has to remain smaller than 1 so $\lambda \leq 1/\sqrt{\rho}$ must hold. 
Under model \eqref{Eq:ModSim}, the motif probabilities can be computed using Proposition \ref{Prop:ProbMotifWProd} where
$$
\xi_h 
= (\sqrt{\rho} \lambda)^h / (h\lambda-h+1).
$$
We considered  graphs of size  $n = 100$ to  $316\;(\simeq 10^{2.5})$
with  log-density $\log_{10}\rho  =  -2, -1.5,  -1$ and  concentration
$\lambda  =  1, 2,  3$  and  $5$. 100  graphs  were  sampled for  each
configuration. 
For each sampled graph, we fitted SBM models with $Q = 1$ to $10$ groups using the VBEM algorithm described above and computed all approximate posterior distributions.

\paragraph{Criteria.}
We used the variational posterior mean $\widehat{w}(u, v) = \widetilde{\Esp}(w(u, v) | \widehat{Q}, \Xbf)$ as an estimate of $W(u, v)$, where $\widehat{Q}$ stands for the maximum \emph{a posteriori} (MAP) estimate of $Q$
$$
\widehat{Q} = \arg\max_Q \widetilde{p}(Q | \Xbf).
$$
Marginalizing over $Q$ (i.e. taking
$\widehat{w}(u, v) = \widetilde{\Esp}(w(u, v) | \Xbf)$) provided similar results in all configurations (not shown). To assess the quality of this estimation of $W$, we computed the root mean squared error ($RMSE$) between its true value and its variational posterior mean, that is
$$
RMSE = \sqrt{\iint \left[ W(u, v) -  \widehat{w}(u, v) \right]^2\dd u \dd v }.
$$
The integral was evaluated on a thin grid over $[0, 1]^2 \times [0, 1]$. \\
As for the motif probability, we considered all motifs $m$ with 2, 3 and 4 nodes. For each of them we computed its probability $\mu(\mbf)$ and we used its variational posterior mean as an estimate: $\widehat{\mu}(\mbf) = \widetilde{\Esp}(\mu(\mbf) |  \widehat{Q}, \Xbf)$. To compare the two, we used the Kullback-Leibler divergence between the corresponding Bernoulli distribution, that is
$$
KL(\mbf) = \mu(\mbf) \log \frac{\mu(\mbf)}{\widehat{\mu}(\mbf)} + (1-\mu(\mbf)) \log \frac{1-\mu(\mbf)}{1-\widehat{\mu}(\mbf)}.
$$

\subsection{Results}

\paragraph{Computational cost.}
First, in order to give some insight into the computational cost of the
proposed methodology, we recorded the running time for the
inference of $W$-graph models in various scenarios. In this section, we set
$\lambda=2$. The results presented  in Table \ref{table:compcost} were
obtained  on  an   Intel  Xeon  CPU  3.07GHz,  a   unique  core  being
used. It appears that estimates are obtained in less than $30$ seconds, even for dense
($\rho=10^{-1}$) networks with $n=316$ nodes. As expected, the running
time is lower for sparse networks, \emph{i.e.} as $\rho$ decreases. 
\begin{table}[ht]
\begin{centering}
\begin{tabular}{c|ccc}
  \hline
 size of the network ($n$) & $\rho=10^{-1}$ & $\rho=10^{-1.5}$ & $\rho=10^{-2}$ \\ 
  \hline
100 & 5.64 s & 5.10 s& 5.20 s\\ 
  147 & 5.95 s& 5.74 s& 5.35 s\\ 
  215 & 8.71 s& 7.85 s& 6.49 s\\ 
  316 & 22.09 s& 19.61 s& 14.47 s\\ 
   \hline
\end{tabular}
\caption{Averaged running time (in seconds) for the $W$-graph model inference procedure, for
  various sizes $n$ of networks and various graph densities $\rho$.} 
\label{table:compcost}
\end{centering}
\end{table}

\begin{figure}[htbp] \centering
  \setlength{\unitlength}{5mm}
  \includegraphics[width=.8\textwidth]{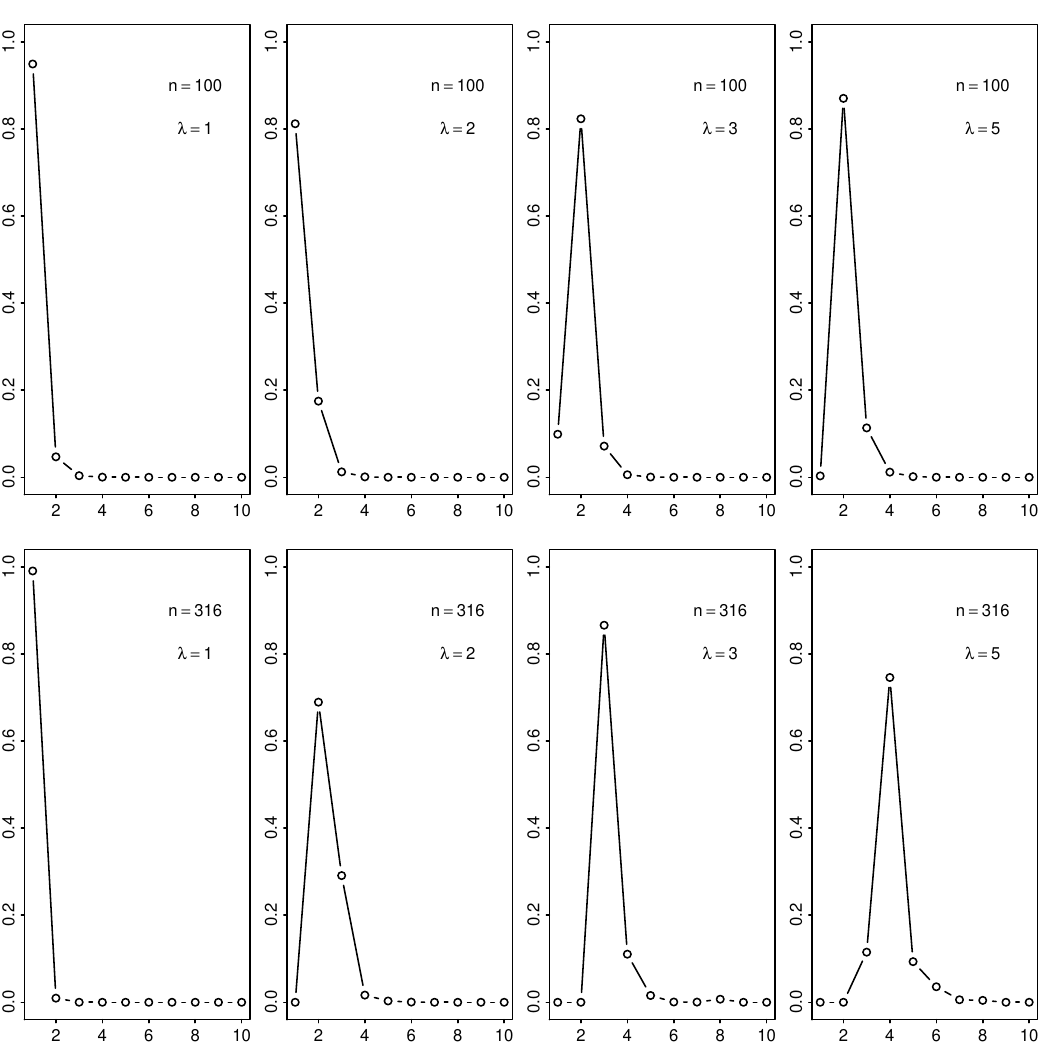}
  \caption{Approximate posterior distribution $\tilde{p}(Q| \Xbf)$ of the number $Q\in\{1,\dots,10\}$ of classes in
    the SBM model,  for $\rho=10^{-1.5}$ and various values  of $n$ as
    well as $\lambda$. \label{Fig:Q}}
 \end{figure}

\paragraph{Model complexity.}
Then, we  studied the (approximate) posterior distribution of $Q$. Figure \ref{Fig:Q}
shows how  the SBM model adapts  to the graphon shape,  using a higher
number of classes as the $W$-graph model becomes more distinct from the
Erd\"os-Renyi model, that is as  $\lambda$ increases. We see that, for
a same non-Erd\"os-Renyi graph ($\lambda > 1$), a more complex SBM can
be  fitted with a  larger graph  size $n$.  We also  see that  for the
Erd\"os-Renyi model ($\lambda=1$), the posterior distribution $Q$ is more concentrated on the true value $Q=1$ when $n$ is larger. The last observation is that all posterior distributions are concentrated around $\widehat{Q}$, resulting in similar results when using the MAP distribution $\widetilde{p}(\cdot | \widehat{Q}, \Xbf)$ or the averaged one $\widetilde{p}(\cdot | \Xbf) = \sum_Q \widetilde{p}(Q | \Xbf) \widetilde{p}(\cdot | \Xbf, Q)$.

\begin{figure}[htbp] \centering
  \setlength{\unitlength}{5mm}
 \includegraphics[width=.8\textwidth]{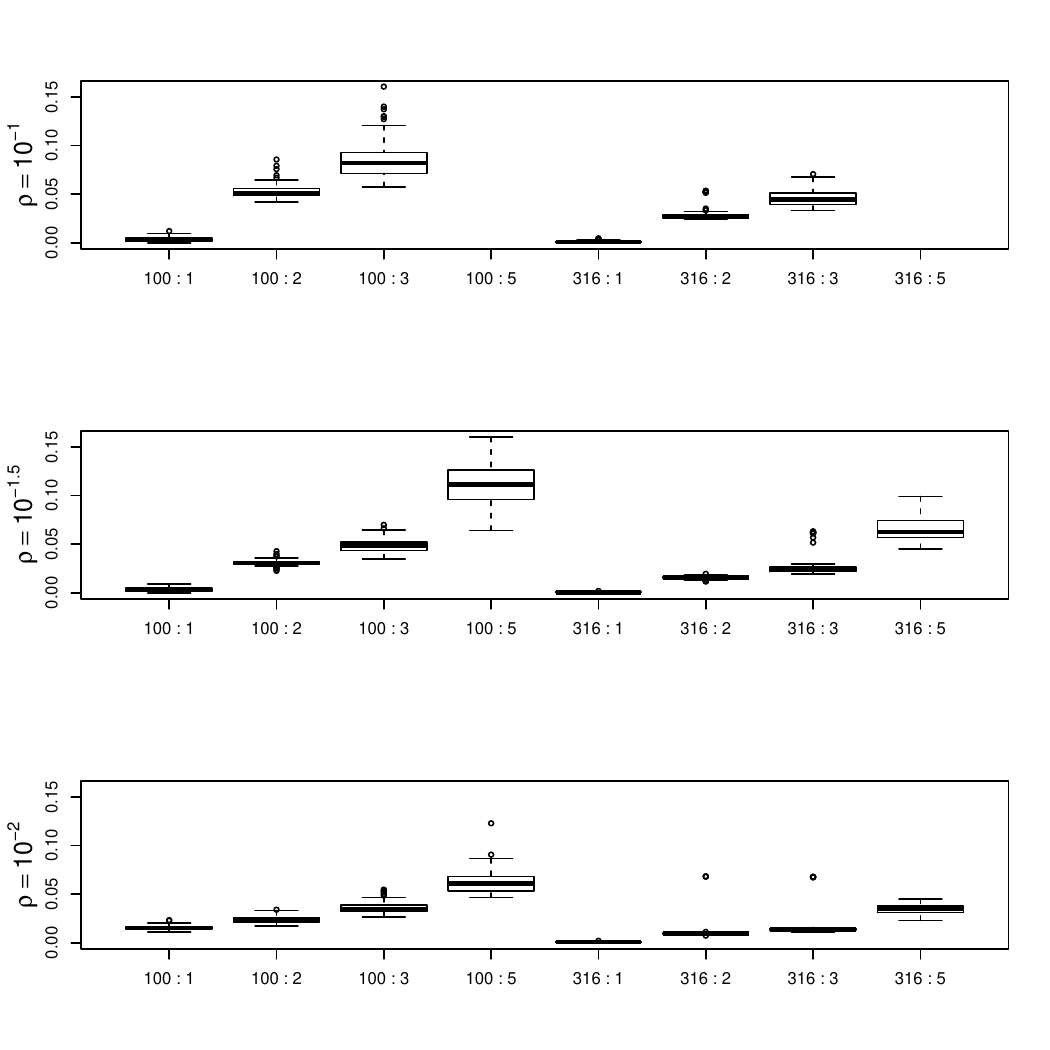}
  \caption{$RMSE$ of  the estimate of  the graphon function  for graph
    density $\rho = 10^{-2}$, $10^{-1.5}$ and $10^{-1}$. $x$-axis: graph size $n$ and shape $\lambda$ labeled as $n : \lambda$. \label{Fig:Graphon}}
 \end{figure}

\paragraph{Estimation of $W$.}
Figure \ref{Fig:Graphon} shows that the RMSE of the estimate is usually below few percent. As expected, the most difficult configurations are imbalanced ($\lambda \geq 3$) medium size ($n=100$) graphs. The RMSE also increases with $\rho$ but this only reflects the fact that $\rho$ is the mean value of $W$, so the error increases with it. However, the relative RMSE ($RMSE/\rho$) actually decreases with $\rho$ (not shown). 

\begin{figure}[htbp] \centering
  \setlength{\unitlength}{5mm}
  \includegraphics[width=.8\textwidth]{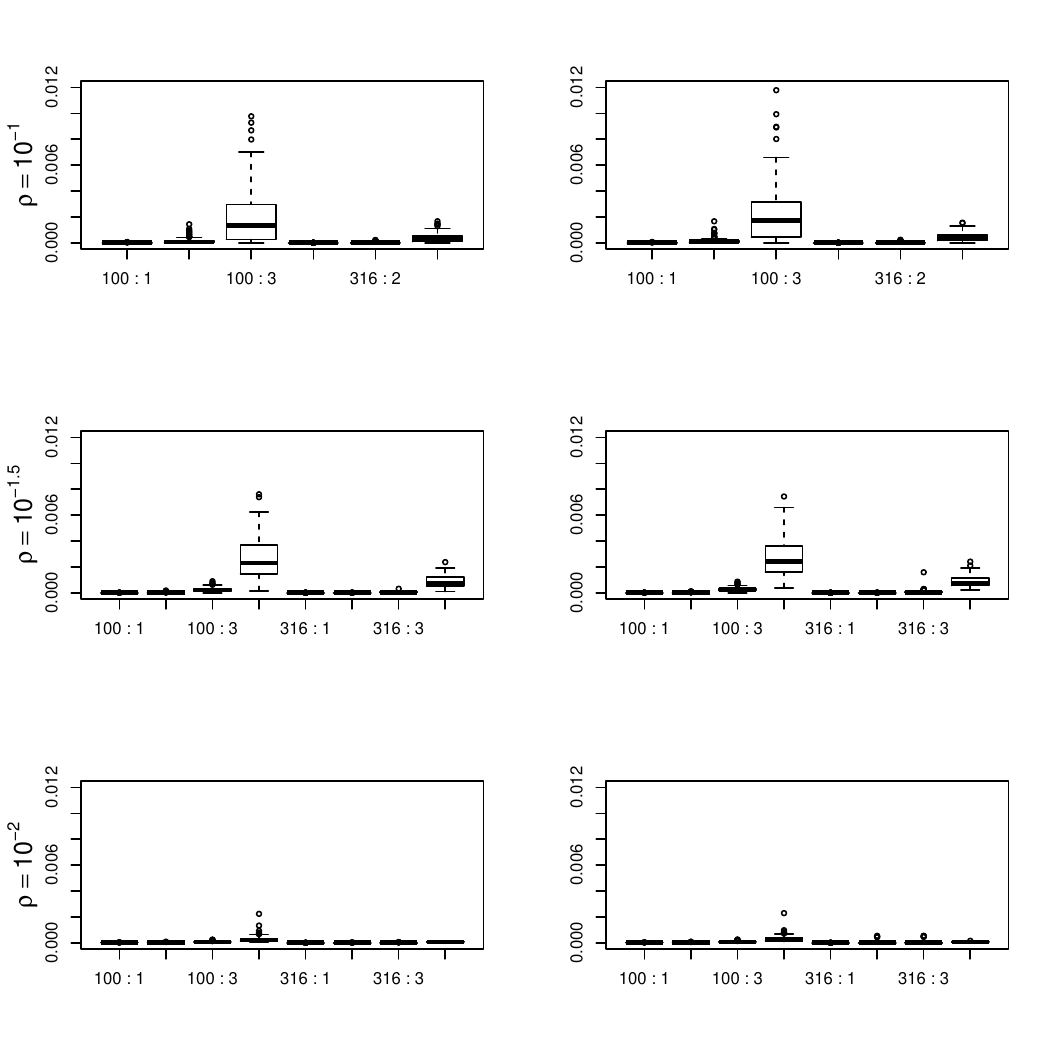}
  \caption{KL divergence between the true and estimated probabilities for the triangle (left) and square (right) motif. Same legend as Figure \ref{Fig:Graphon}. \label{Fig:Square}}
 \end{figure}

\paragraph{Motif probability.}
We then turned to the motifs probabilities and the results are given in
Figure \ref{Fig:Square}. We remind that these quantities are invariant and identifiable in the $W$-graph, as opposed to the graphon function $W$. The estimation turns out to be very good, even for very imbalanced shape ($\lambda=5$) as long as the graph if large ($n = 10^{1.5}$) and dense no too dense ($\rho = 10^{-2}$).



%% file: Examples.tex
\section{French political blogosphere} \label{Sec:Experiments}

As in \cite{articlelatouche2011}, we consider a subset of the French
political blogosphere network. The network is made of 196 vertices
connected by 2864 edges. It was built from a single day snapshot of
political blogs automatically extracted on 14th october 2006 and
manually classified by the `Observatoire Pr\'{e}sidentiel'' project
\citep{articlezanghi2008}. Nodes correspond to hostnames and there is
an edge between two nodes if there is a known hyperlink from one
hostname to the other. The four main political parties which are present
in the data set are the UMP (french ``republican''), liberal party
(supporters of economic-liberalism), UDF (``moderate''
party), and PS (french ``democrat''). We run the VBEM algorithm on the data set for $Q \in \{1,\dots,
20\}$ using the R package \emph{mixer}.



\begin{figure}[htbp] \centering
 \setlength{\unitlength}{5mm}
 \includegraphics[scale=0.8]{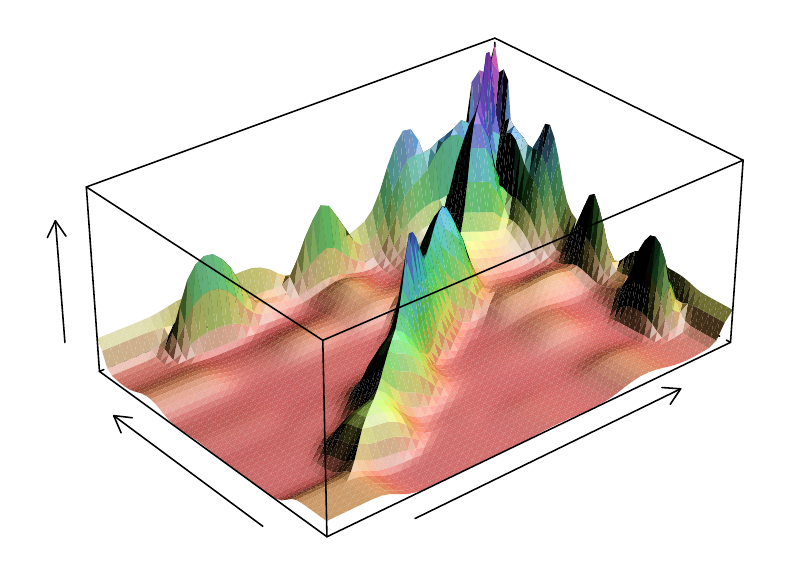}
 \caption{Graphon function of the French political blogosphere
 network  estimated  using the  estimated  posterior  mean derived  in
 Section  \ref{ssec:postW}.}
 \label{fig:resultsBlog}
\end{figure}

\paragraph{Graphon function.}
The  graphon function  estimated using  the model  averaging approach
we   proposed  in   Section  \ref{ssec:postW}   is  given   in  Figure
\ref{fig:resultsBlog}.  For  this   network,  we  emphasize  that  the
estimated  posterior distribution  of  the number  $Q$  of classes  is
highly concentrated around $Q^{*}=12$. As in the preceding section, all prior parameters were set to 1 to induce uniform priors. 

\begin{figure}[htbp] \centering
 \setlength{\unitlength}{5mm}
 \includegraphics[width=9.5cm, height=9.5cm]{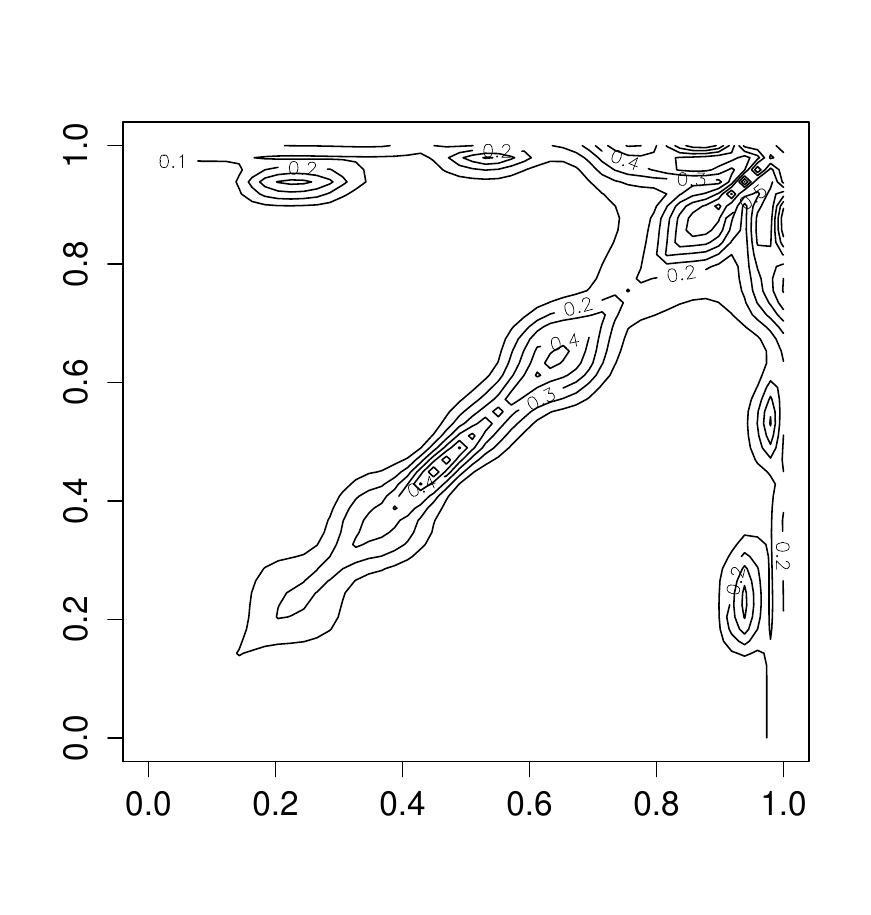}
 \caption{Contour plot of the French political blogosphere
 network graphon function estimated  using the  estimated  posterior  mean derived  in
 Section  \ref{ssec:postW}.}
 \label{fig:resultsBlogContour}
\end{figure}

First, we notice that high  connectivity regions appear in two series of hills,  one along the diagonal  and one parallel  to the $y$-axis, close to $x=1$ (we recall that this function is symmetric).
The series of hills on the diagonal each corresponds to a specific political party. In terms of connection patterns, the diagonal structure reveals that blogs of the given community more likely connect to blogs of the same community. Moreover, we emphasize that  the plateau as the very bottom
 left hand side of the graphon function represents blogs from various political parties, from the left wing to the right
wing, having very weak  connection profiles. Conversely, the series of
hills  parallel to  the axes  correspond to  blogs, and  in particular
blogs of political analysts, having strong connections with the different political parties. Because of the identifiability rule which makes the degree $D(x) = \int W(x, y) \dd y$ increasing, this region also corresponds to nodes with highest degree. From a global point of view, this region of the graphon plays a critical role as it ensures the connectivity of the whole network.
Yet, a closer look at the contour plot given in Figure \ref{fig:resultsBlogContour} shows thats the modes of these hills all have a $x$ coordinate close to 1 but also have very different $y$ coordinates, which reveals that some of these blogs have themselves preferential connections with specific political parties.

\begin{figure}
\begin{center}
\begin{tabular}{ccrrrr}
  \hline
  & Motifs & $N_{obs}$ & $\Esp[N(\mbf)]$ & $\sqrt{\mathbb{V}(N(\mbf))}$ & Sd. diff. \\ 
  \hline
  1 & \begin{tabular}{c}\includegraphics[width=0.12\textwidth, height=0.12\textwidth, clip=]{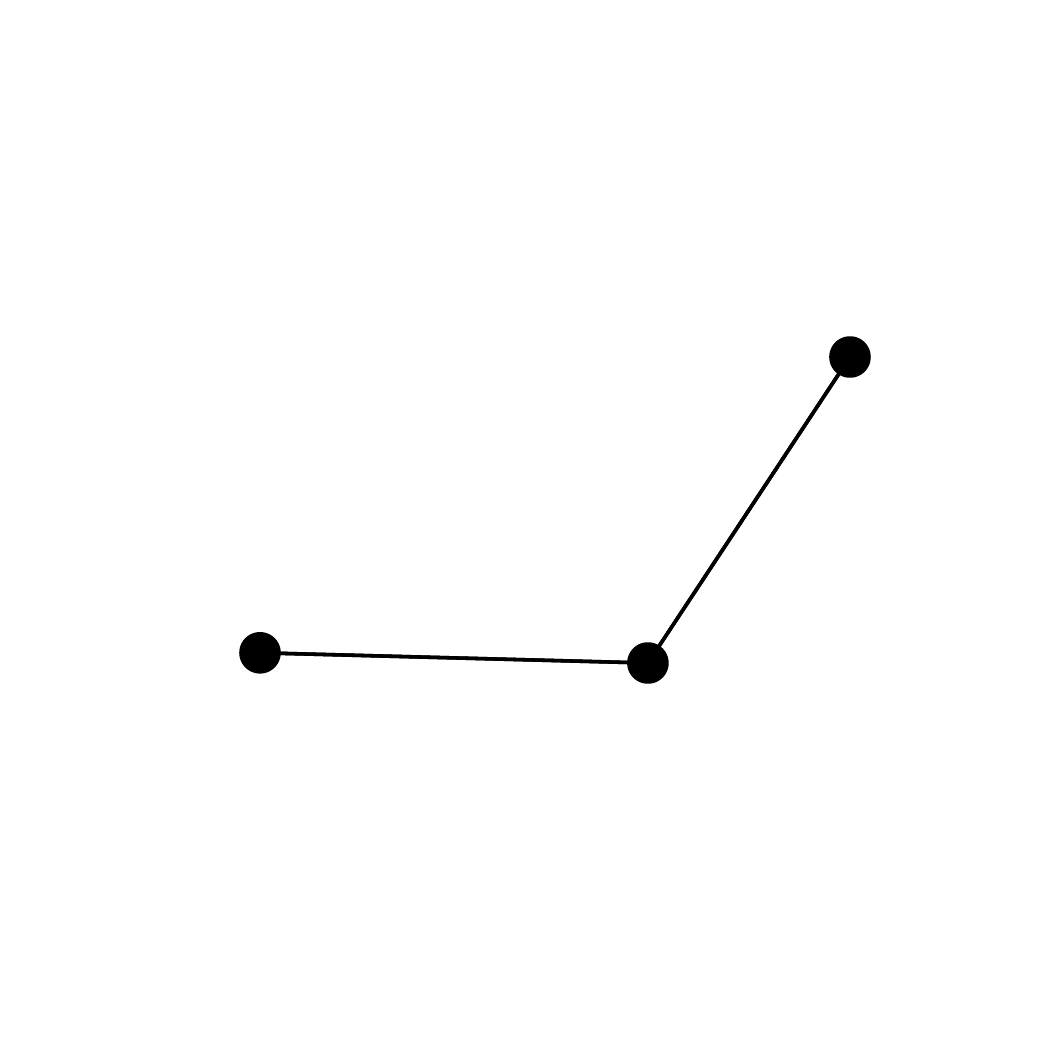} \end{tabular}& 29715 & 39722.11 & 8259.28 & -1.21\\ 
  2 & \begin{tabular}{c}\includegraphics[width=0.12\textwidth, height=0.12\textwidth, clip=]{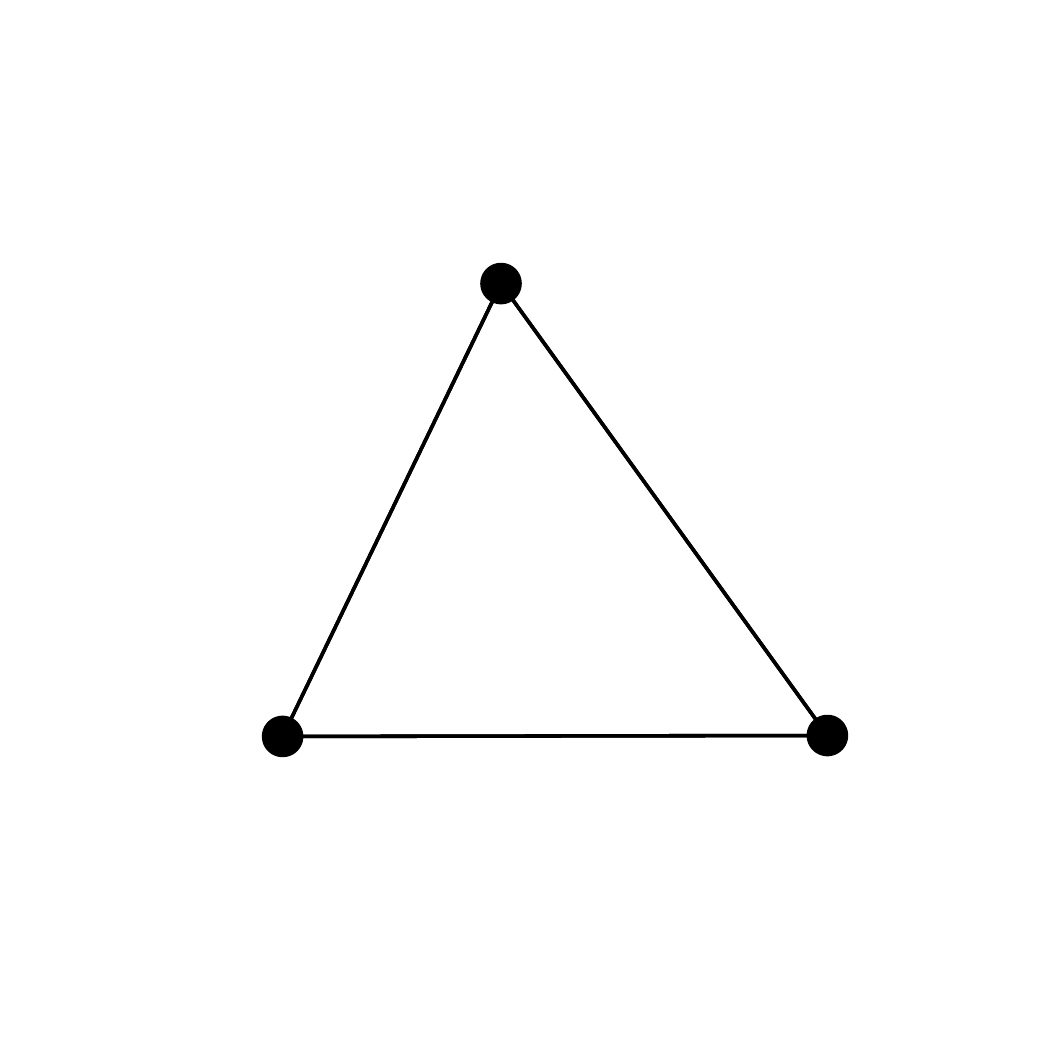} \end{tabular} & 3821 & 4512.51 & 1276.13 & -0.54\\ 
  3 & \begin{tabular}{c}\includegraphics[width=0.12\textwidth, height=0.12\textwidth, clip=]{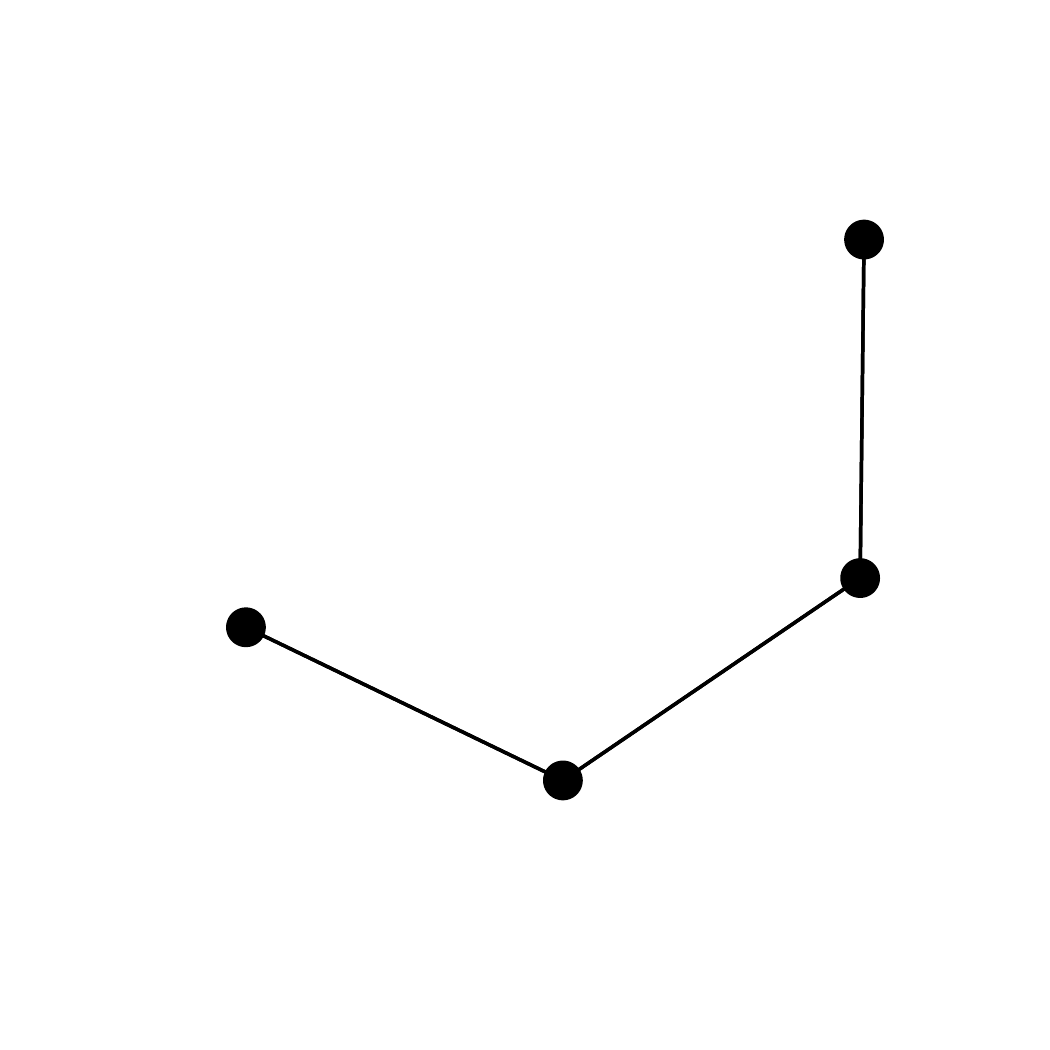} \end{tabular} & 608708 & 968364.08 & 336800.24 & -1.07 \\ 
  4 & \begin{tabular}{c}\includegraphics[width=0.12\textwidth, height=0.12\textwidth, clip=]{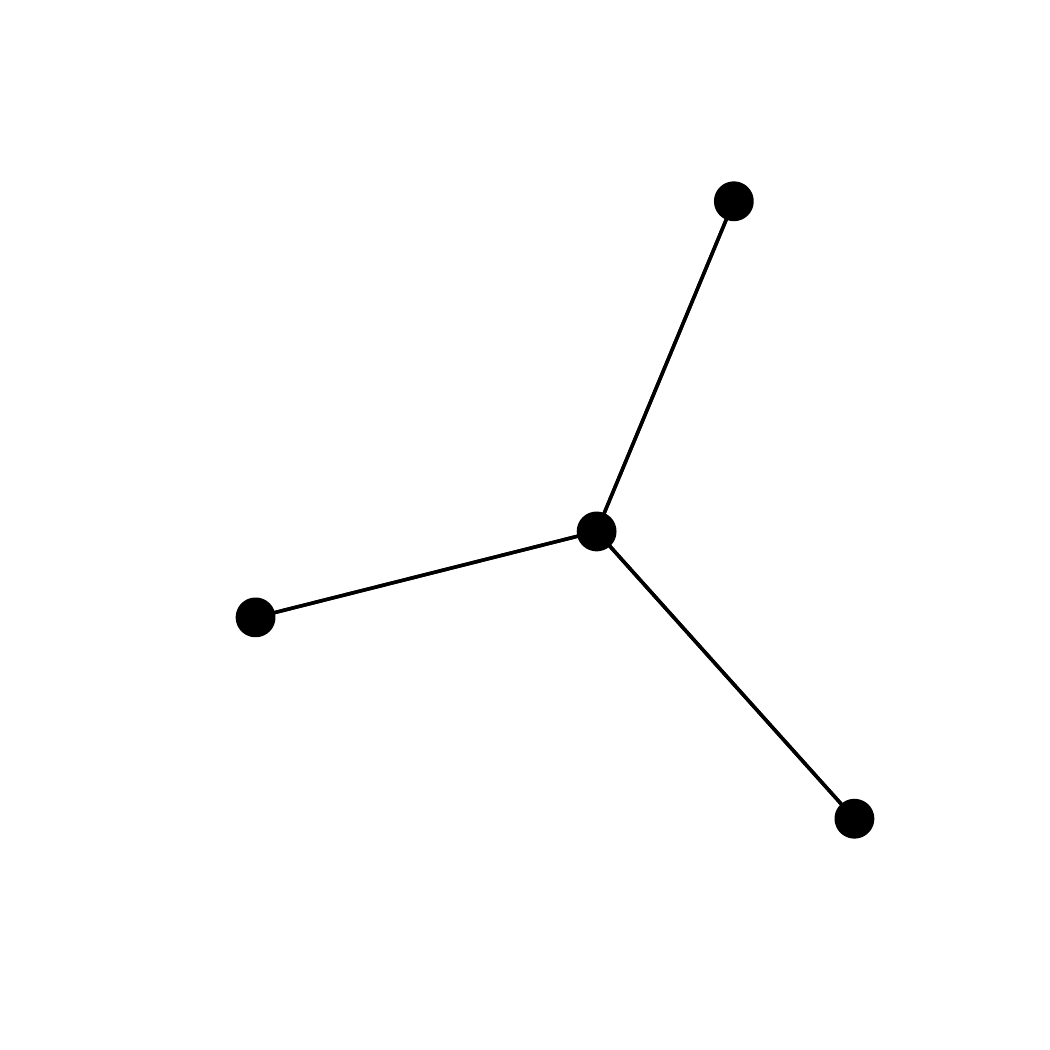} \end{tabular} & 279771 & 428867.52 & 153962.02 & -0.97 \\ 
  5 & \begin{tabular}{c}\includegraphics[width=0.12\textwidth, height=0.12\textwidth, clip=]{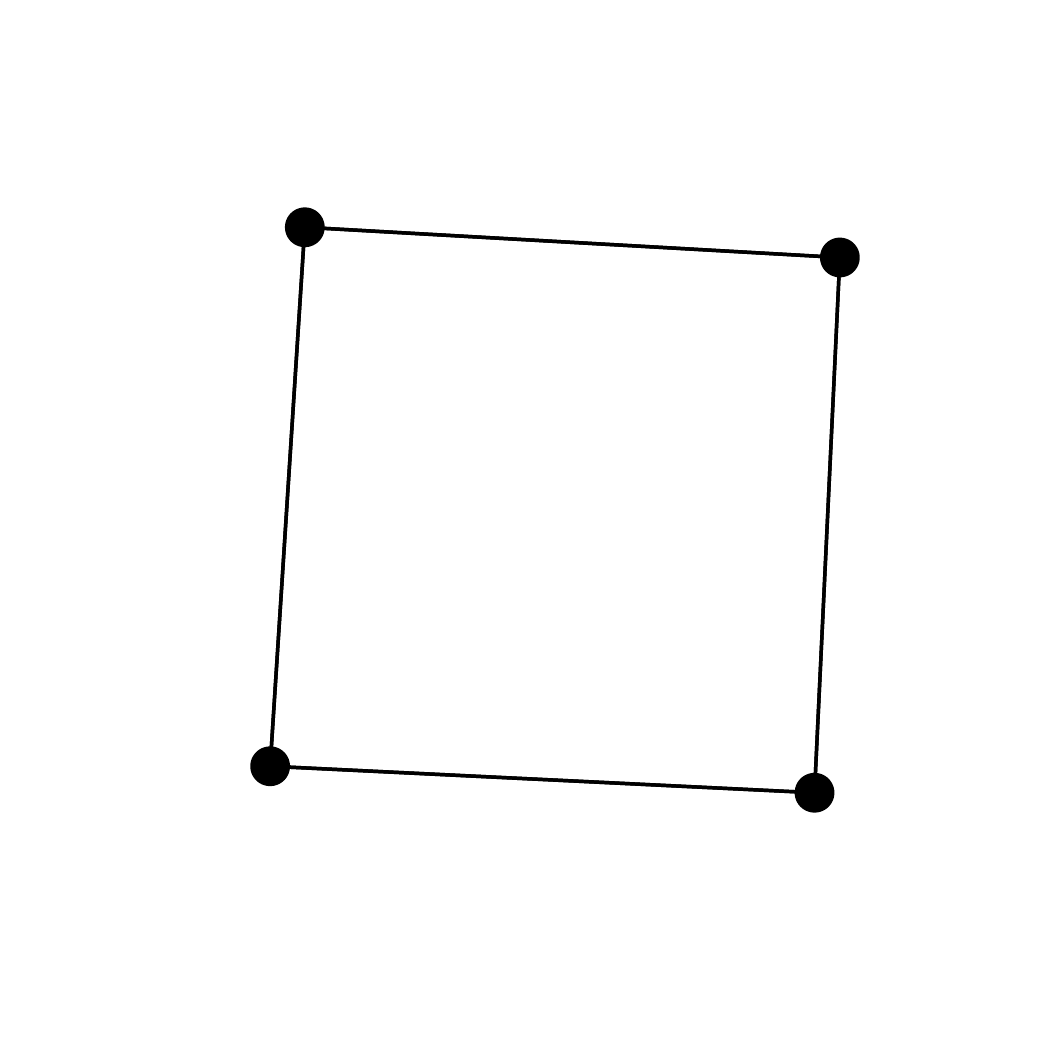} \end{tabular} & 47415 & 74533.94 & 35075.09 & -0.77 \\ 
  6 & \begin{tabular}{c}\includegraphics[width=0.12\textwidth, height=0.12\textwidth, clip=]{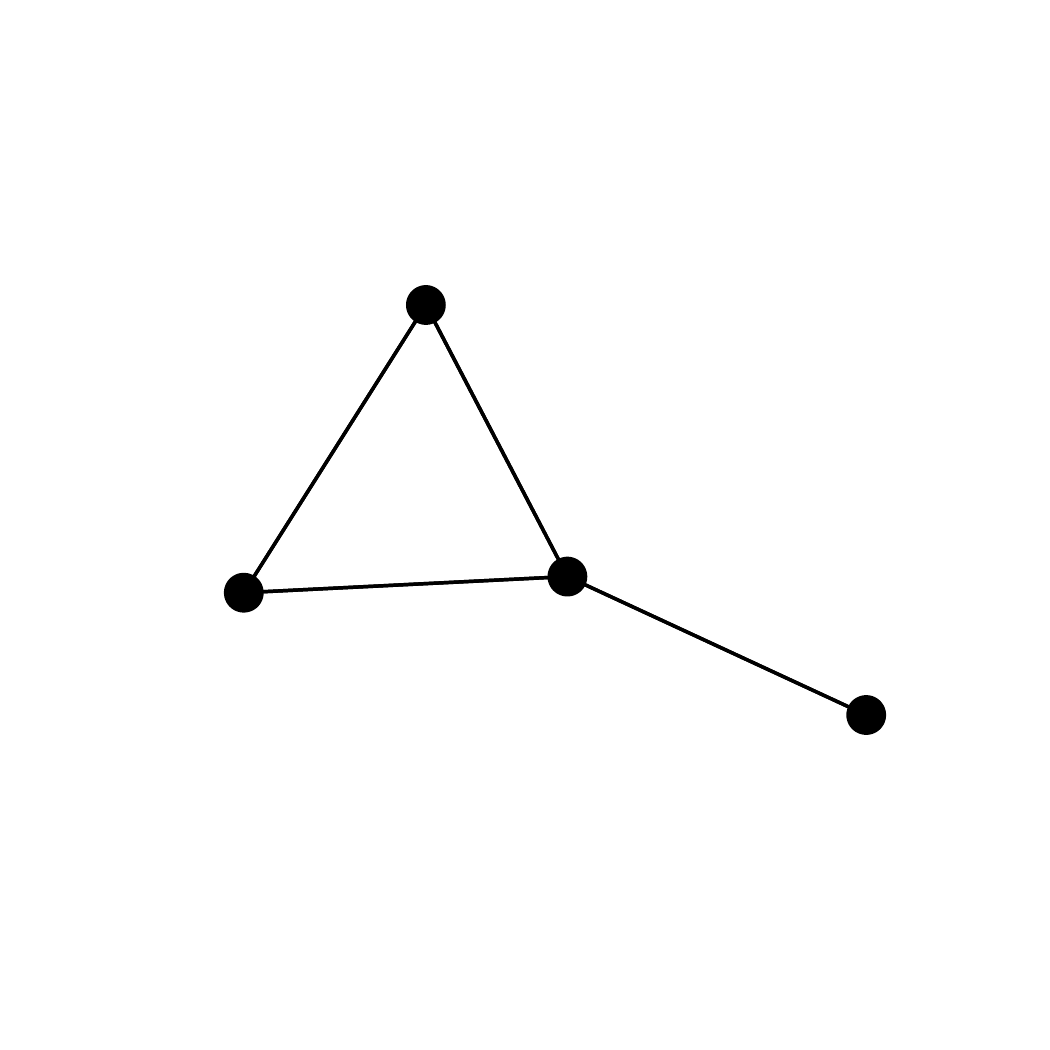} \end{tabular} & 270497 & 397053.82 & 177049.01 & -0.71 \\ 
  7 & \begin{tabular}{c}\includegraphics[width=0.12\textwidth, height=0.12\textwidth, clip=]{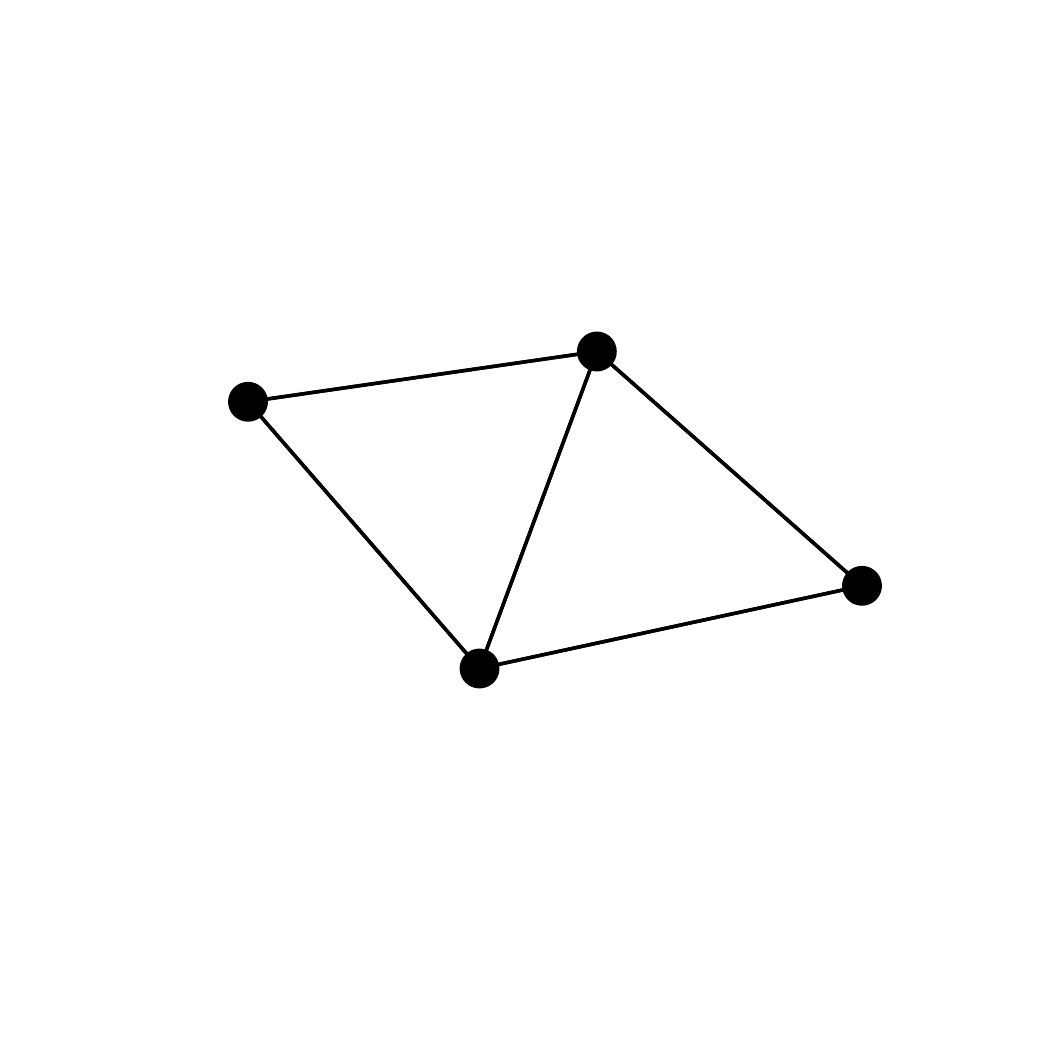} \end{tabular}  & 62071 & 87849.83 & 47407.22 & -0.54 \\ 
  8 & \begin{tabular}{c}\includegraphics[width=0.12\textwidth, height=0.12\textwidth, clip=]{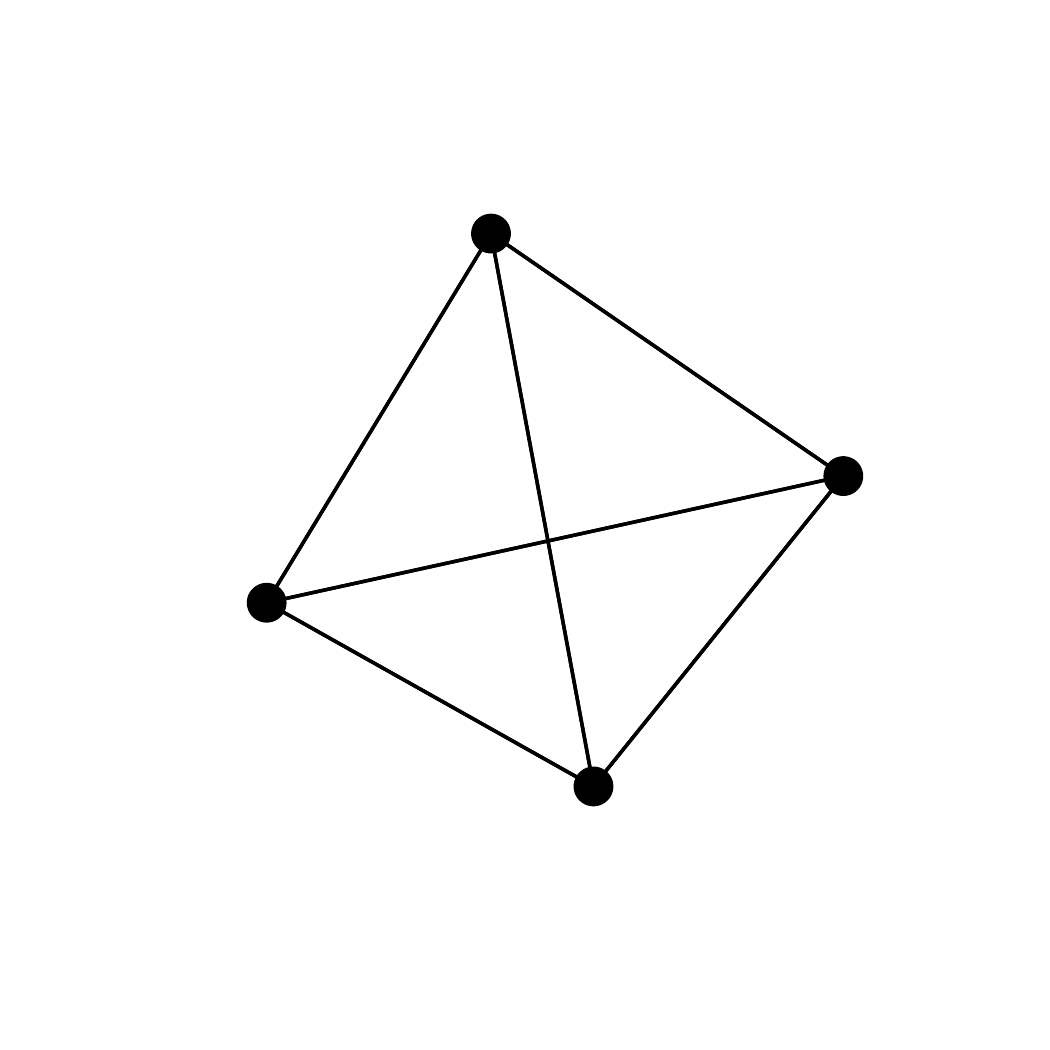} \end{tabular}   & 6523 & 8818.95 & 5385.87 & -0.43 \\ 
  \hline
\end{tabular}
\caption{$N_{obs}$  true  counts  of    motifs  in  the  French
  political  blogosphere   network;  $\Esp[N(\mbf)]$  estimated  means;
  $\sqrt{\mathbb{V}(N(\mbf))}$  estimated  standard  errors; Sd. diff. (standardized difference $(N_{obs} - \Esp[N(\mbf)]) / \sqrt{\mathbb{V}(N(\mbf))}$.}
\label{tab:motifs}
\end{center}
\end{figure}

\paragraph{Motif frequency.}
To complete the analysis  of the French political blogosphere network,
we  also  computed the  estimated  mean  and  standard
deviation  of the motif count $N(\mbf)$,
for various  motifs. As pointed out  in Section \ref{ssection:testing},
both  quantities involve  motif occurrence  probabilities $\mu(\cdot)$
for which we provide estimators in Section \ref{ssection:motifprob}, using the variational inference
procedure.   Our  results  are
summarized  in Table  \ref{tab:motifs}.  \\
First, its  appears that  the
three  motifs which are  mainly present  in the  network are  motif 3
(4-edges path), motif 4 (3-branch star), and motif 6 (triangle plus an
edge). However, none  of the motifs are seen  as unexpectedly frequent
motifs. Indeed, we found that their  observed counts are less than $1.5$
standard deviation away for their means under the $W$-graph model. This means that the $W$-graph model, estimated using the
variational  approach, explains  reasonably well  the presence  of the
motifs  in the  network, such  that no  counts $N_{obs}$  are  seen as
unexpected.  This tends to illustrate the goodness of fit of the estimated
$W$-graph model.  

In the same vein, we would like to stress that random graph models in
social  sciences  often  consider  specific parameters  to  explain  the
presence of triangles in networks (see for instance \cite{gr07}).  The additional parameter dedicated to triangles aims at accounting for the 'friends of my friends are my friends' effect. Conversely, a $W$-graph model focuses on modeling  edges between pairs of nodes.  Triangles are not specifically modeled and only result from the  construction of edges between  triads. Interestingly, for the social  network of blogs
we considered, we  found that the observed count is less than one standard deviation away from its mean under the $W$-graph model. Again,
this tends  to show that the  presence of triangles in  the network is
sufficiently explained by the estimated $W$-graph model and that the 'friends of my friends' effects is accounted for by the latent position of the actors.

%% file: Conclusion.tex
\section{Conclusion}

We  considered the  $W$-graph model  which generalizes
most of  the random  graph models commonly  used in the  literature to
extract  knowledge  from  network  topologies. The  model  is  defined
through  a  
graphon function $W$ which  has to  be inferred in  practice while working  on real
data.  \\
To this aim, we
relied on  a variational approximation  procedure originally developed
for the SBM model that can be seen as a $W$-graph model with
blockwise  constant graphon function. Then,  we showed  how the  approximate posterior distribution over the SBM model parameters (including the number of blocks) could be integrated out analytically to  obtain an estimate  of the posterior  distribution of
the graphon  function.  \\
Using the same approach,  we  derived  the  approximate  posterior  mean  of  motifs
frequencies. We propose to use this expected frequencies under the $W$-graph as a goodness-of-fit criterion for this model. In the blogosphere application, we conclude that the most widely studied motifs display a frequency that is consistent with the $W$-graph model.

%% file: App-InfW.tex
\subsection{Inference of the function $W$} \label{App:InferenceW}

\paragraph{Proof of Proposition \ref{Prop:PosteriorW}.}
The first part is straightforward, based on a conditioning of the
binnings of $u$ and $v$
\begin{eqnarray*}
\widetilde{p}(w(u, v)|\Xbf, Q) & = & \widetilde{p}(\pi_{C(u), C(v)}|\Xbf,
Q) \\ & = & \sum_{q \leq \ell} \widetilde{p}(\pi_{q, \ell} | \Xbf, Q,
C(u) = q, C(v) = \ell) \widetilde{\Pr}\{C(u) = q, C(v) = \ell | \Xbf,
Q\} \\ & = & \sum_{q \leq \ell} b(w; \eta_{q, \ell}, \zeta_{q, \ell})
\widetilde{\Pr}\{C(u) = q, C(v) = \ell | \Xbf, Q\}.
\end{eqnarray*}
We are now left with the calculation of 
\begin{eqnarray*}
\widetilde{\Pr}\{C(u) = q, C(v) = \ell | \Xbf, Q\} & = &
\widetilde{\Pr}\{\sigma_{q-1} < u < \sigma_q, \sigma_{\ell-1} < v <
\sigma_\ell | \Xbf, Q\} \\
& = & 
F_{{q-1}, {\ell-1}}(u, v; \abf) - F_{{q}, {\ell-1}}(u, v; \abf) - F_{{q-1}, {\ell}}(u, v; \abf) \\
&& + F_{{q}, {\ell}}(u, v; \abf)
\end{eqnarray*} 
where 
\begin{itemize}
\item   $\abf,  \boldsymbol{\eta}$   and   $\boldsymbol{\zeta}$  are
the parameters of the variational Bayes
posterior distributions;
\item $b(\cdot; \boldsymbol{\eta}, \boldsymbol{\zeta})$ stands for the pdf of the Beta
distribution $\Beta(\boldsymbol{\eta}, \boldsymbol{\zeta})$;
\item $F_{q, \ell}(u, v; \abf)$ denotes the joint cdf of $(\sigma_q,
\sigma_\ell)$, as defined in \eqref{Eq:CumProp}, when $\alphabf$ has a Dirichlet distribution
$\Dir(\abf)$.
\end{itemize}
The last argument comes from \cite{GoS10} who give explicit recursions
to compute the uni- and bi-variate cdf for the Dirichlet $\Dir(\abf)$,
denoted $ G_{q}(u; \abf)$ and $G_{q, \ell}(u, v; \abf)$ respectively.
\\ 
Reminding that the approximate variational posterior of $\alphabf$ is
$\Dir(\abf)$ and using a simple property of the Dirichlet
distribution
$$
(\alphabf) \sim \Dir(\abf) 
\quad \Rightarrow \quad 
\left(\sum_{j=1}^q \alpha_j, \sum_{j=q+1}^\ell \alpha_j, \sum_{j=\ell+1}^Q
\alpha_j\right) \sim \Dir\left(\sum_{j=1}^q a_j, \sum_{j=q+1}^\ell a_j,
\sum_{j=\ell+1}^Q a_j\right),
$$ 
the calculation of $F_{q, \ell}(u, v)$ follows as
\begin{eqnarray*}
  F_{q, \ell}(u, v) & = & \widetilde{\Pr}\{\sigma_q < u, \sigma_\ell <
  v | \Xbf, Q\} \\
  & = & \widetilde{\Pr}\{\sigma_q < u, 1 - \sigma_\ell > 1 - v | \Xbf, Q\} \\
  & = & \widetilde{\Pr}\{\sigma_q < u | \Xbf, Q\} - \Pr\{\sigma_q < u,
  \sigma_\ell < 1 - v | \Xbf, Q\} \\
  & = & G_1(u; [s_q, s_\ell-s_q, s_Q-s_\ell]) - G_{1, 3}(u, 1-v; [s_q,
    s_\ell-s_q, s_Q-s_\ell]),
\end{eqnarray*}
where the $(s_q)$ are the cumulated parameters: $s_q = \sum_{j=1}^q
a_j$.
\proofend

%% file: App-Motifs.tex
\subsection{Motif probability} \label{App:Motifs}

\begin{figure}[h]
  \begin{center}
    \begin{tabular}{lcccc}
      \begin{tabular}{l}
        motif $m$
      \end{tabular}
      & \begin{tabular}{c} 
          \includegraphics[width=.12\textwidth, height=.12\textwidth, clip=]{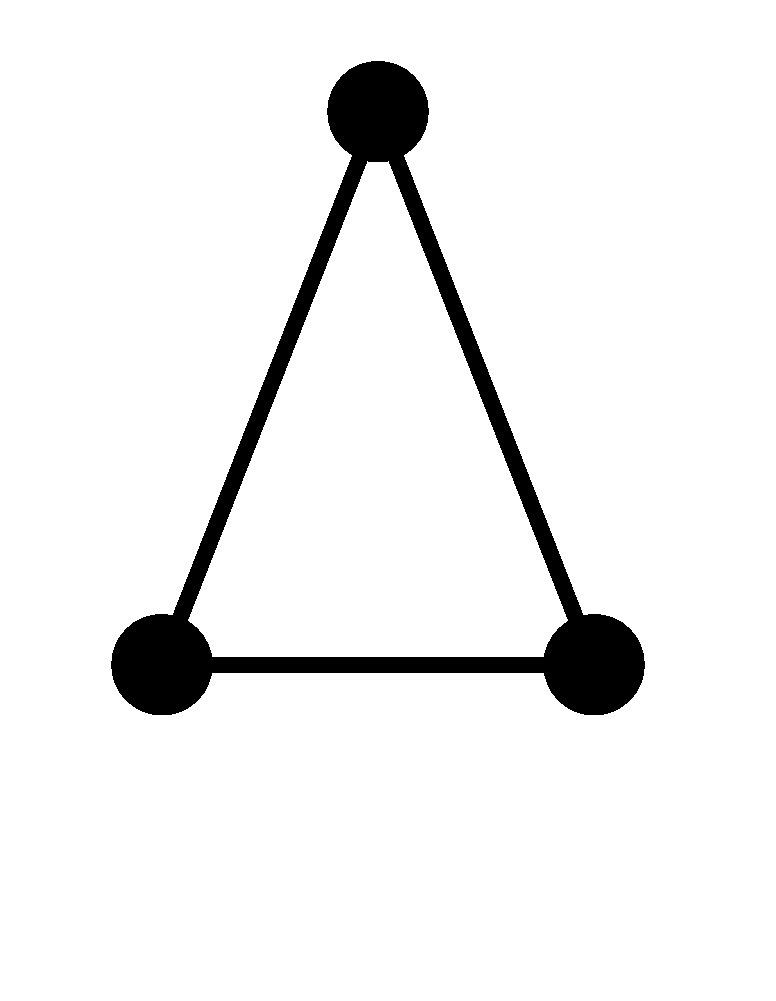}
        \end{tabular}
      & \begin{tabular}{c} 
          \includegraphics[width=.12\textwidth, height=.12\textwidth, clip=]{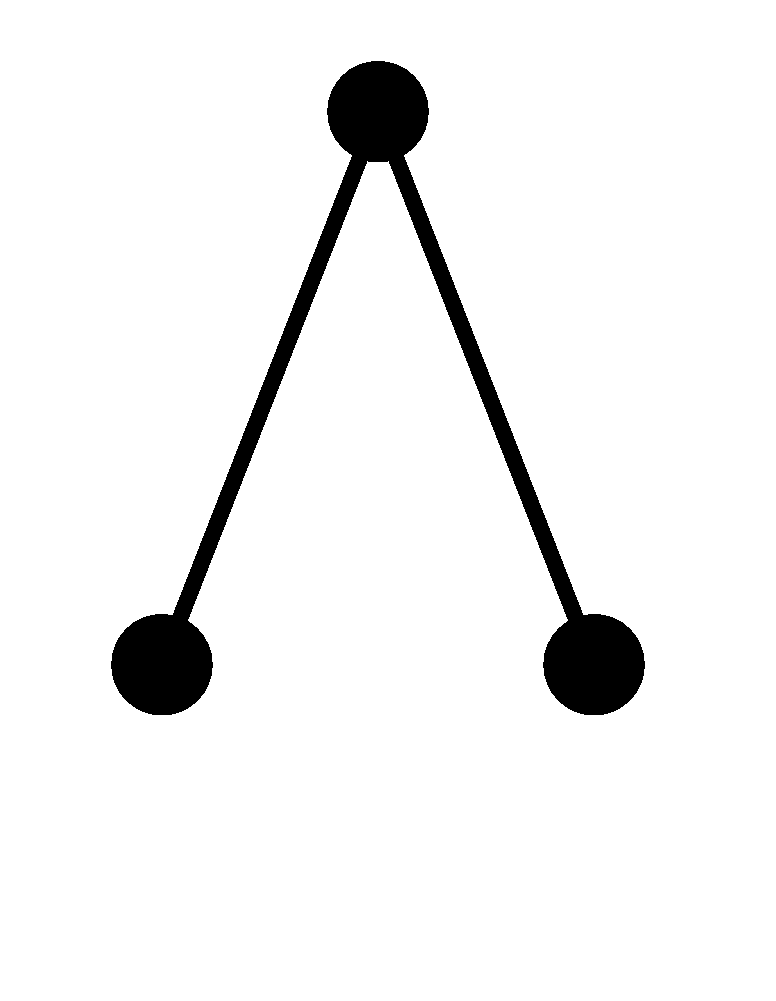}
        \end{tabular}
      & \begin{tabular}{c} 
          \includegraphics[width=.12\textwidth, height=.12\textwidth, clip=]{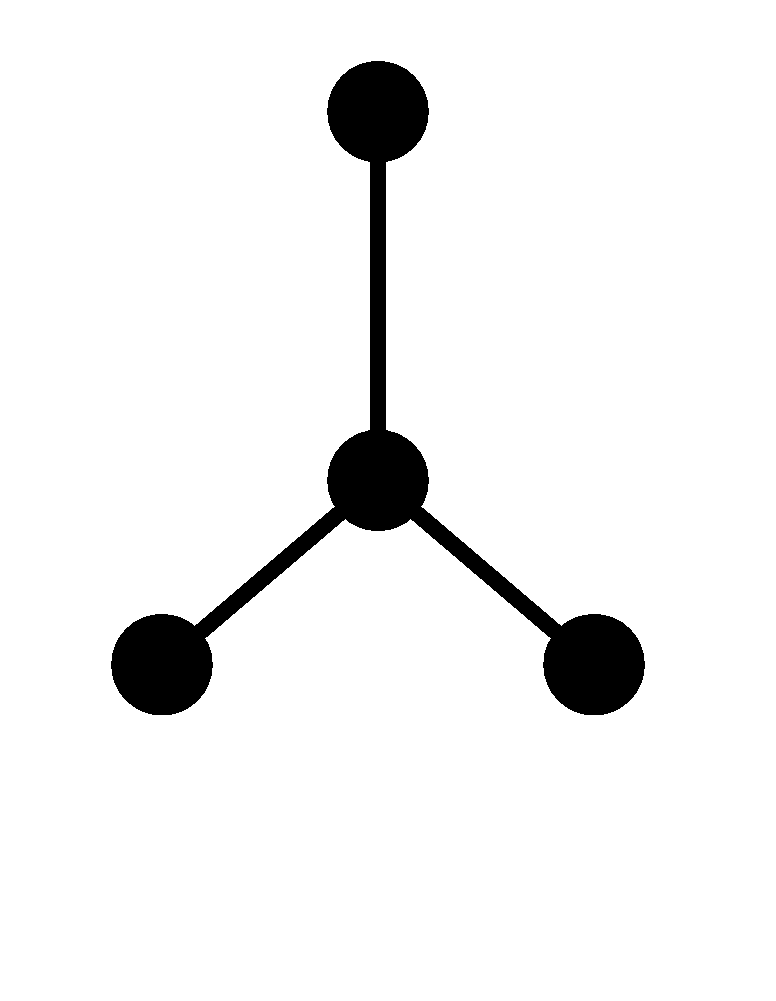}
        \end{tabular}
      & \begin{tabular}{c} 
          \includegraphics[width=.12\textwidth, height=.12\textwidth, clip=]{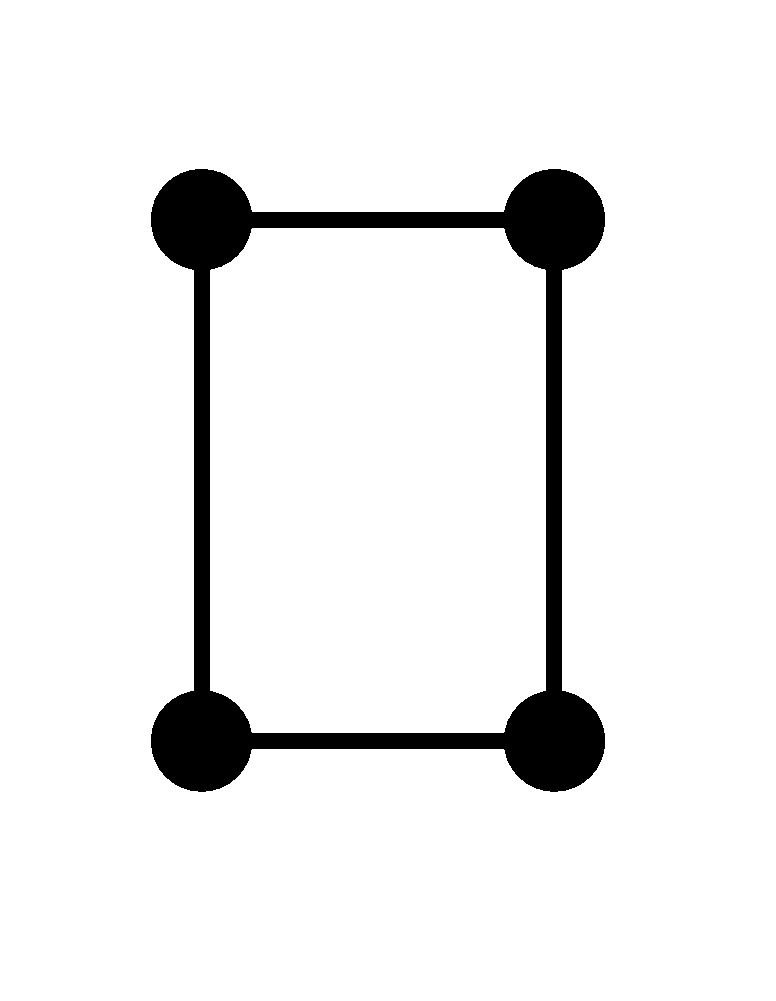}
        \end{tabular}
      \\
      \begin{tabular}{l}
        adjacency \\
        matrix $\mbf$
      \end{tabular}
      &
      $\displaystyle{
        \left[\begin{array}{ccc}
            0 & 1 & 1 \\
            1 & 0 & 1 \\
            1 & 1 & 0
          \end{array}\right]
        }$
      &
      $\displaystyle{
        \left[\begin{array}{ccc}
            0 & 1 & 0 \\
            1 & 0 & 1 \\
            0 & 1 & 0
          \end{array}\right]
      }$
      &
      $\displaystyle{
        \left[\begin{array}{cccc}
            0 & 1 & 1 & 1 \\
            1 & 0 & 0 & 0 \\
            1 & 0 & 0 & 0 \\
            1 & 0 & 0 & 0 
          \end{array}\right]
      }$
      &
      $\displaystyle{
        \left[\begin{array}{cccc}
            0 & 1 & 0 & 1 \\
            1 & 0 & 1 & 0 \\
            0 & 1 & 0 & 1 \\
            1 & 0 & 1 & 0 
          \end{array}\right]
      }$
    \end{tabular}
    \caption{Adjacency matrix $\mbf$ for four typical
      motifs. \label{Fig:Motifs}}
  \end{center}
\end{figure}

\paragraph{Proof of Proposition \ref{Prop:PosterioMumSBM}.}
We directly write the approximate variational expectation
\begin{equation*} \label{eq:occuProb2}
  \begin{aligned}
    \widetilde{\Esp}[\mu(\mbf) | \Xbf, Q] & = \int \int
    \Esp[\mu(\mbf)|\alphabf, \pibf] \widetilde{p}(\alphabf,
    \pibf|{\Xbf, Q})\,\dd\alphabf\,\dd\pibf \\
    &= \int \int \left\{\sum_{\cbf}
    \Esp[\mu(\mbf) |\cbf,\pibf] 
    p(\cbf|\alphabf)\right\}\widetilde{p}(\alphabf,
    \pibf|{\Xbf, Q})\,\dd\alphabf\,\dd\pibf,
  \end{aligned}
\end{equation*}
where 
$$ 
p(\cbf|\alphabf) = \prod_{1 \leq a \leq k} p(c_a|\alphabf) 
= \prod_{1 \leq a \leq k}\prod_{1 \leq q \leq Q}\alpha_{q}^{\Ibb\{c_a = q\}} 
= \prod_{1 \leq q \leq Q} \alpha_{q}^{n_{q}^{\cbf}}.
$$
Furthermore, we have
\begin{equation*} \label{eq:occuProb2}
  \begin{aligned}
    \Esp[\mu(\mbf) |\cbf,\pibf]     
    &= \Pr\left\{\prod_{1 \leq a<b \leq k}X_{{a}{b}}^{m_{ab}} = 1| \cbf, \pibf \right\} 
    = \prod_{1 \leq a<b \leq k}\Pr\left\{X_{{a}{b}} = 1 | c_a, c_b,
    \pibf \right\}^{m_{ab}} \\    
    &= \prod_{1 \leq a<b \leq k}\prod_{1 \leq q , \ell \leq Q}
    \pi_{q\ell}^{\Ibb\{c_a=q\}\Ibb\{c_b=\ell\}m_{ab}} \\    
    &=  \prod_{1 \leq q < \ell \leq Q}\prod_{a \neq
      b}\pi_{q\ell}^{\Ibb\{c_a=q\}\Ibb\{c_b=\ell\}m_{ab}}\prod_{1 \leq q \leq Q}\prod_{1
      \leq a<b \leq k}\pi_{qq}^{\Ibb\{c_a=q\}\Ibb\{c_b=q\}m_{ab}} 
    = \prod_{1 \leq q \leq \ell \leq Q} \pi_{q\ell}^{\eta_{q\ell}^{\cbf}},
  \end{aligned}
\end{equation*}
so we end up with
\begin{eqnarray*} 
    \widetilde{\Esp}[\mu(\mbf) | \Xbf, Q]
    & = & \int \int \sum_{\cbf}  \prod_{1 \leq q \leq \ell \leq Q}
    \pi_{q\ell}^{\eta_{q\ell}^{\cbf}}  
    \prod_{1 \leq q \leq Q}\alpha_{q}^{n_{q}^{\cbf}} 
    \widetilde{p}(\alphabf,\pibf| {Q})\,\dd\alphabf \,\dd\pibf \\
    & = & \int \int \sum_{\cbf}\prod_{1 \leq q \leq \ell \leq Q}
    \pi_{q\ell}^{\eta_{q\ell}^{\cbf}}   \prod_{1 \leq q \leq Q}
    \alpha_{q}^{n_{q}^{\cbf}}   \prod_{1 \leq q \leq \ell \leq Q}
    \frac{\Gamma(\eta_{q\ell}+\zeta_{q\ell})}{\Gamma(\eta_{q\ell})\Gamma(\zeta_{q\ell})}
    \pi_{q\ell}^{\eta_{q\ell}-1} (1-\pi_{q\ell})^{\zeta_{q\ell}-1}
     \\
    & & \qquad \frac{\Gamma(\sum_{1 \leq q \leq Q}n_{q})}{\prod_{1 \leq q \leq Q}\Gamma(n_{q})}
    \prod_{1 \leq q \leq Q}\alpha_{q}^{n_{q}-1}\,\dd\alphabf \,\dd\pibf \\
    & = & \sum_{\cbf} \prod_{1 \leq q \leq \ell \leq Q}
    \frac{\Gamma(\eta_{q\ell}+\zeta_{q\ell})}{\Gamma(\eta_{q\ell})\Gamma(\zeta_{q\ell})}
    \int \pi_{q\ell}^{\eta_{q\ell} + n_{q\ell}^{\cbf}-1}
    (1-\pi_{q\ell})^{\zeta_{q\ell}-1} \,\dd \pi_{q\ell}  \\
    & & \qquad \frac{\Gamma(\sum_{1 \leq q \leq Q}n_{q})}{\prod_{1 \leq q \leq Q}\Gamma(n_{q})}\prod_{1 \leq q \leq Q}\int
    \alpha_{q}^{n_{q}+n_{q}^{\cbf}-1} \,\dd \alphabf_{q} \\
    & = &\sum_{\cbf}  \prod_{1 \leq q \leq \ell \leq Q}
    \frac{\Gamma(\eta_{q\ell}+\zeta_{q\ell})}{\Gamma(\eta_{q\ell})\Gamma(\zeta_{q\ell})}
    \frac{\Gamma(\eta_{q\ell}+\eta_{q\ell}^{\cbf})\Gamma(\zeta_{q\ell})}{\Gamma(\eta_{q\ell}+\eta_{q\ell}^{\cbf}+\zeta_{q\ell})}    
    \frac{\Gamma(\sum_{1 \leq q \leq Q}n_{q})}{\prod_{1 \leq q \leq Q}\Gamma(n_{q})}\frac{\prod_{1 \leq q \leq Q}\Gamma(n_{q}+n_{q}^{\cbf})}{\Gamma
      \sum_{1 \leq q \leq Q} (n_{q}+n_{q}^{\cbf}) },
\end{eqnarray*}
and the proof is completed. \proofend

\paragraph{Proof of Proposition \ref{Prop:ProbMotifWProd}.}
Because the $Z_i$'s are uniformly distributed over $[0; 1]$, we have
\begin{eqnarray*}
 \mu(\mbf) & = & \Pr\{Y(i_1, \dots i_k; m) = 1\} \\
  & = & \idotsint \Pr\left\{\prod_{1 \leq a < b \leq k} X_{i_a i_b}^{m_{ab}} = 1 | Z_{i_1}=z_1, \dots Z_{i_k} = z_k\right\} \dd z_1 \dots \dd z_k \\
  & = & \idotsint \prod_{1 \leq a < b \leq k} [w(z_a)w(z_b)]^{m_{ab}}\dd z_1 \dots \dd z_k \\
  & = & \idotsint \prod_{1 \leq a \leq k} w(z_a)^{m_{a+}} \dd z_1 \dots \dd z_k 
  \quad = \quad \prod_{1 \leq a \leq k} \int w(z)^{m_{a+}} \dd z.
\end{eqnarray*}
\proofend